\documentclass[a4paper,fleqn,usenatbib]{mnras}

\usepackage[normalem]{ulem}

\usepackage{newtxtext,newtxmath}

\usepackage[T1]{fontenc}
\usepackage{ae,aecompl}

\usepackage{graphicx}	
\usepackage{amsmath}	
\usepackage{amssymb}	

\title[PATHOS\,I: LCs of 47\,Tuc]{A PSF-based Approach to TESS High
  quality data Of Stellar clusters (PATHOS) - I. Search for exoplanets
  and variable stars in the field of 47\,Tuc.  }

\author[D.\ Nardiello et al]{D.\ Nardiello$^{1,2}$\thanks{E-mail: domenico.nardiello@unipd.it},
  L.\ Borsato$^{1,2}$,
  G.\ Piotto$^{1,2}$,
  L.\ S.\ Colombo$^{1}$,
  E.\ E.\ Manthopoulou$^{1}$,
  \newauthor
  L.\ R.\ Bedin$^{2}$,
  V.\ Granata$^{1}$,
  G.\ Lacedelli$^{1}$,
  M.\ Libralato$^{3}$,
  L.\ Malavolta$^{4}$,
  M.\ Montalto$^{1,2}$,
  \newauthor
  V. Nascimbeni$^{2,1}$ \\
$^{1}$Dipartimento di Fisica e Astronomia ``Galileo Galilei'', Universit\`a di Padova, Vicolo dell'Osservatorio 3, IT-35122, Padova, Italy  \\
$^{2}$Istituto Nazionale di Astrofisica - Osservatorio Astronomico di Padova, Vicolo dell'Osservatorio 5, IT-35122, Padova, Italy \\
$^{3}$Space Telescope Science Institute, 3800 San Martin Drive, Baltimore, MD 21218, USA \\
$^{4}$Istituto Nazionale di Astrofisica - Osservatorio Astronomico di Catania, Via S. Sofia 78, IT-95123, Catania, Italy \\
}

\date{Accepted 2019 October 4. Received 2019 September 19; in original form 2019 June 12}

\pubyear{2019}

\begin{document}
\label{firstpage}
\pagerange{\pageref{firstpage}--\pageref{lastpage}}
\maketitle

\begin{abstract}

  The {\it TESS} mission will survey $\sim$85\,\% of the sky, giving
  us the opportunity of extracting high-precision light curves of
  millions of stars, including stellar cluster members. In this work,
  we present our project ``A PSF-based Approach to TESS High quality
  data Of Stellar clusters'' (PATHOS), aimed at searching and
  characterise candidate exoplanets and variable stars in stellar
  clusters using our innovative method for the extraction of
  high-precision light curves of stars located in crowded
  environments.  Our technique of light-curve extraction involves the
  use of empirical Point Spread Functions (PSFs), an input catalogue
  and neighbour-subtraction. The PSF-based approach allows us to
  minimise the dilution effects in crowded environments and to extract
  high-precision photometry for stars in the faint regime ($G>13$).

  For this pilot project, we extracted, corrected, and analysed the
  light curves of 16641 stars located in a dense region centred on the
  globular cluster 47\,Tuc. We were able to reach the {\it TESS}
  magnitude $T \sim 16.5$ with a photometric precision of $\sim 1\%$
  on the 6.5-hour timescale; in the bright regime we were able to
  detect transits with depth of $\sim$34 parts per million. We
  searched for variables and candidate transiting exoplanets. Our
  pipeline detected one planetary candidate orbiting a main sequence
  star in the Galactic field. We analysed the period-luminosity
  distribution for red-giant stars of 47\,Tuc and the eclipsing
  binaries in the field. Light curves are uploaded on the Mikulski
  Archive for Space Telescopes under the project PATHOS.

\end{abstract}




\section{Introduction}
In the last two decades, more than 4000
exoplanets\footnote{\url{http://exoplanets.eu/}} have been discovered
and confirmed using spectroscopic and photometric studies. More than
75\%  exoplanets have been found using the transit method; a
strong contribution to this research area came from pioneering photometric
surveys, carried out both with ground-based (e.g. SuperWASP,
\citealt{2006PASP..118.1407P}) and space telescopes (e.g. {\it CoRoT},
\citealt{2006cosp...36.3749B}).  In last years, the
{\it Kepler} main mission (\citealt{2010Sci...327..977B})
 and the reinvented {\it Kepler/K2} mission
(\citealt{2014PASP..126..398H}), allowed the astronomical
community to develop new techniques to extract and analyse  high precision light curves of
stars in many Galactic fields and to find thousands of new
exoplanets and variable stars (see, e.g., \citealt{2014PASP..126..948V,2016MNRAS.456.2260A,2016A&A...594A.100B,2016MNRAS.459.2408A,2016MNRAS.456.1137L}).

\begin{figure}
  \includegraphics[width=0.5 \textwidth]{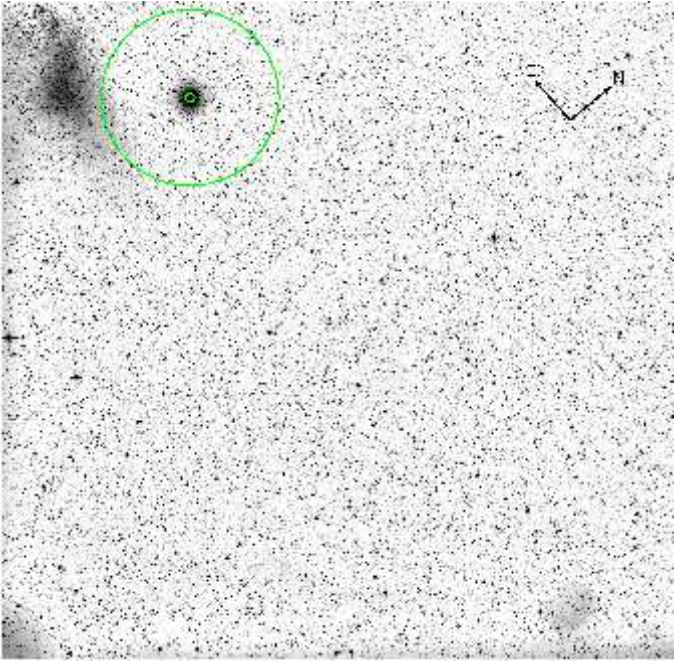}
  \caption{Field of view covered by CCD 2 of Camera 3 in
    Sector-1. The figure shows the stacked image obtained by combining
    1282 \textit{TESS} exposures. The green annulus are the inner and outer bounds of the region
    containing the stars for which we extracted the light curves
    ($0.075 \le R \le 1.5$\,degree). \label{fig1}}
\end{figure}

Nevertheless, to date only a few confirmed planets have been found to
orbit stars in stellar clusters (e.g., \citealt{2012ApJ...756L..33Q,
  2014ApJ...787...27Q, 2014A&A...561L...9B, 2017A&A...603A..85B,
  2016A&A...588A.118M, 2018AJ....155....4M}).  Stellar clusters give
us an unique opportunity to understand how exoplanets have formed
and evolved. Properties of stars in stellar clusters  (e.g.,
mass, radius, age, and chemical composition) can generally be more
reliably measured than for many field stars. For this reason, the
discovery and characterisation of exoplanets orbiting stars in
 stellar clusters allow us to correlate planet and hosting
star properties. Moreover, because stellar clusters' ages span
over a wide range that goes from few Myrs up to ten Gyrs, the analysis of
exoplanets in different stellar clusters allows us to understand how
exoplanets have formed and evolved. In addition, the different chemical
properties of the stellar clusters give us the opportunity to understand how the
environment has affected the exoplanets' life.

Despite all these advantages, a few exoplanets have been found in
stellar clusters because only few clusters have been studied with {\it
  Kepler} and/or spectroscopic surveys; indeed, searches for
exoplanets orbiting stellar cluster members have been hampered by
observational difficulties because of the crowding. For this reason,
appropriate techniques for the extraction of high-precision light
curves of stars in dense environments are mandatory. Recently, studies based on the
extraction of light curves from {\it Kepler} data of open clusters
using the Point Spread Function (PSF) approach
(\citealt{2016MNRAS.456.1137L}) or the differential image analysis
(\citealt{2017PASP..129d4501S}) have allowed us to obtain high precision
photometric time series for very faint stars located in crowded
environments.

In the last year, {\it Kepler} passed the baton to the {\it Transiting
  Exoplanet Survey Satellite} ({\it TESS},
\citealt{2015JATIS...1a4003R}), a NASA all-sky survey mission aimed at
searching exoplanets around bright stars. During its two-year mission,
{\it TESS} will observe about 85\,\% of the sky in 27-day sectors,
each one covering $24 \times 96$ degree$^2$ and most of them
containing hundreds of stellar clusters. In every sector, about
20\,000 stars are observed in short-cadence mode (2-minute). In
addition, {\it TESS} produces onboard stacked images of the entire
field of view, with a cadence of 30-minute (Full-Frame Images, FFIs).
The FFIs allow us to extract light curves for all objects with
$V\lesssim 17$ that fall within the field of view of each sector,
giving us not only the possibility to widen the search of exoplanets
to a huge number of stars, but also to produce results in other fields, like
asteroseismology, analysis of Solar System objects
(\citealt{2018PASP..130k4503P}), variable stars, supernovae
(e.g. \citealt{2019arXiv190308665V}), and other Galactic and
extragalactic sources.

\begin{figure*}
  \includegraphics[width=0.9\textwidth,bb= 15 17 504 580]{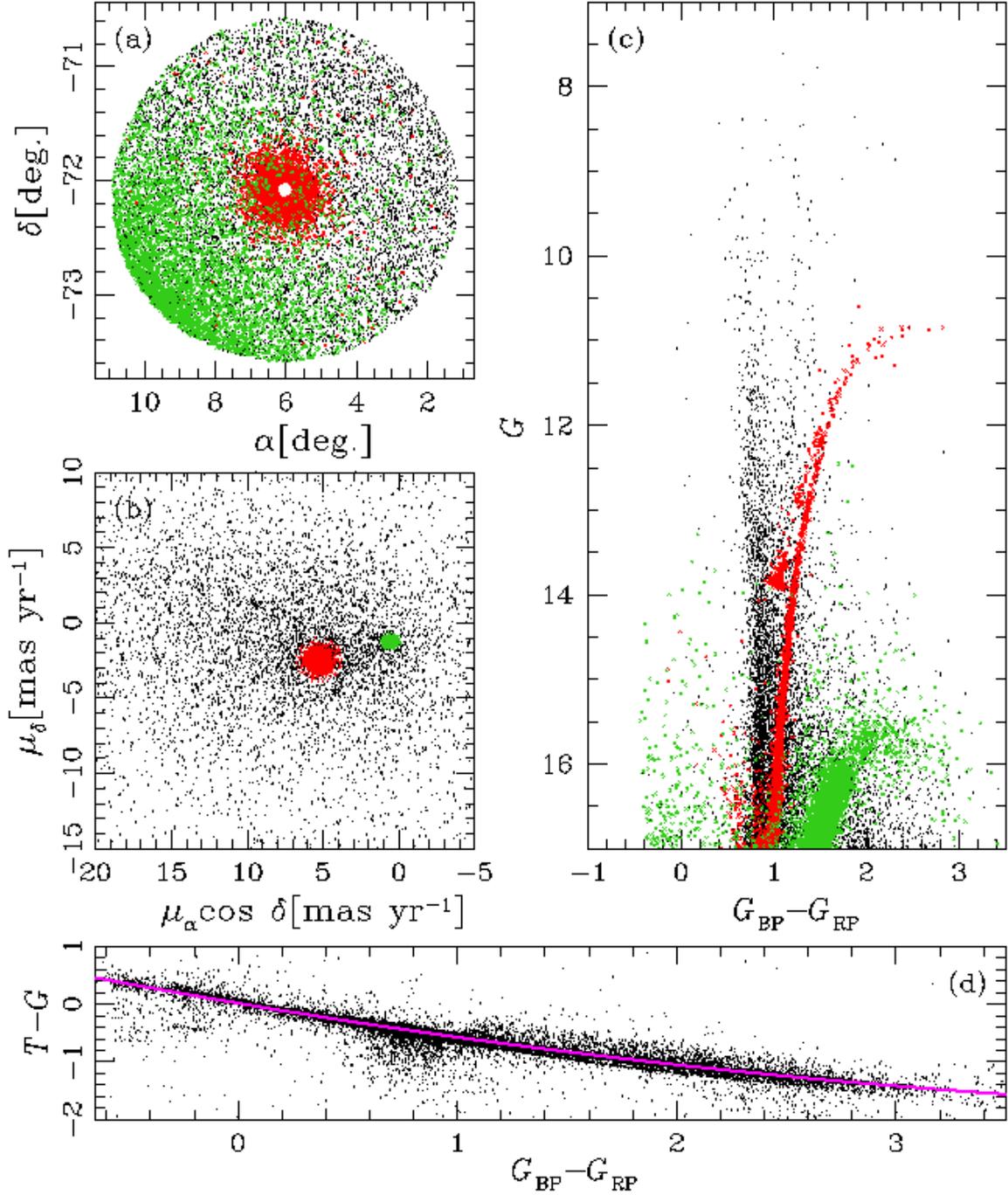}
  \caption{Overview of the Gaia\,DR2 input catalogue used in this work: panel
    (a) plots the $(\alpha,\delta)$-coordinates of the sources in the
    catalogue; panel (b) shows the absolute proper motion distribution
    for all the stars in the catalogue; panel (c) gives the $G$ versus
    $G_{\rm BP}-G_{\rm RP}$ CMD for the stars in the catalogue;
    panel (d) shows the
    $T-G$ versus $G_{\rm BP}-G_{\rm RP}$ distribution for the stars
    observed by \textit{TESS} in the first seven sectors. In magenta the
    relation adopted to transform \textit{TESS} magnitudes in
    \textit{Gaia} magnitudes. Red and green points in panels (a)-(c) are
    stars with high probability to be 47\,Tuc and SMC members, respectively. \label{fig2}}
\end{figure*}

In this work we present our project ``A PSF-based Approach to TESS
High quality data Of Stellar clusters'' (PATHOS), aimed at extracting
high precision light curves of sources in stellar clusters in order to
find candidate exoplanets and variable stars. For this project we
apply our expertise on PSF photometry and astrometry on images of
crowded fields. In fact, because of the sampling of {\it TESS} ($\sim
21$\,arcsec/pixel), also sparse stellar clusters appear to be crowded
on {\it TESS} images. In our previous works
(e.g. \citealt{2015MNRAS.447.3536N,2016MNRAS.456.1137L}) we
demonstrated that by using as input a high-resolution astro-photometric
catalogue and empirical PSFs, we are able to measure the flux of each
target star in the catalogue with extreme accuracy, after subtracting
all its neighbour stars. Using the PSF-based approach we are able to:
(1) extract the light curves of stars in crowded regions; (2)
minimise the light-contamination effects due to neighbour stars, improving the photometric precision;
(3) extract light curves of faint stars, increasing the number of
analysable objects.

In the present work we apply for the first time the PSF-based approach
to {\it TESS} data, in order to extract high-precision light curves of
sources in a very crowded field centred on the globular cluster
NGC\,104 (47\,Tuc). In Section~\ref{sec:ffi} we describe our pipeline
for the extraction of raw light curves and their correction. The
description of the finding, vetting, and modelling of candidate
exoplanets is reported in
Section~\ref{sec:candexo}. Section~\ref{sec:varsta} is a description
of the procedure used to find and analyse variable
stars. In Section~\ref{sec:compare} we compare our pipeline
  with the most advanced, publicly available pipeline for the
  extraction of light curves from FFIs. The public data release
discussion and a summary of the present work are reported in
Section~\ref{sec:dr} and \ref{sec:conc}, respectively.

\begin{figure*}
  \includegraphics[width=0.95\textwidth,bb= 26 440 584 693]{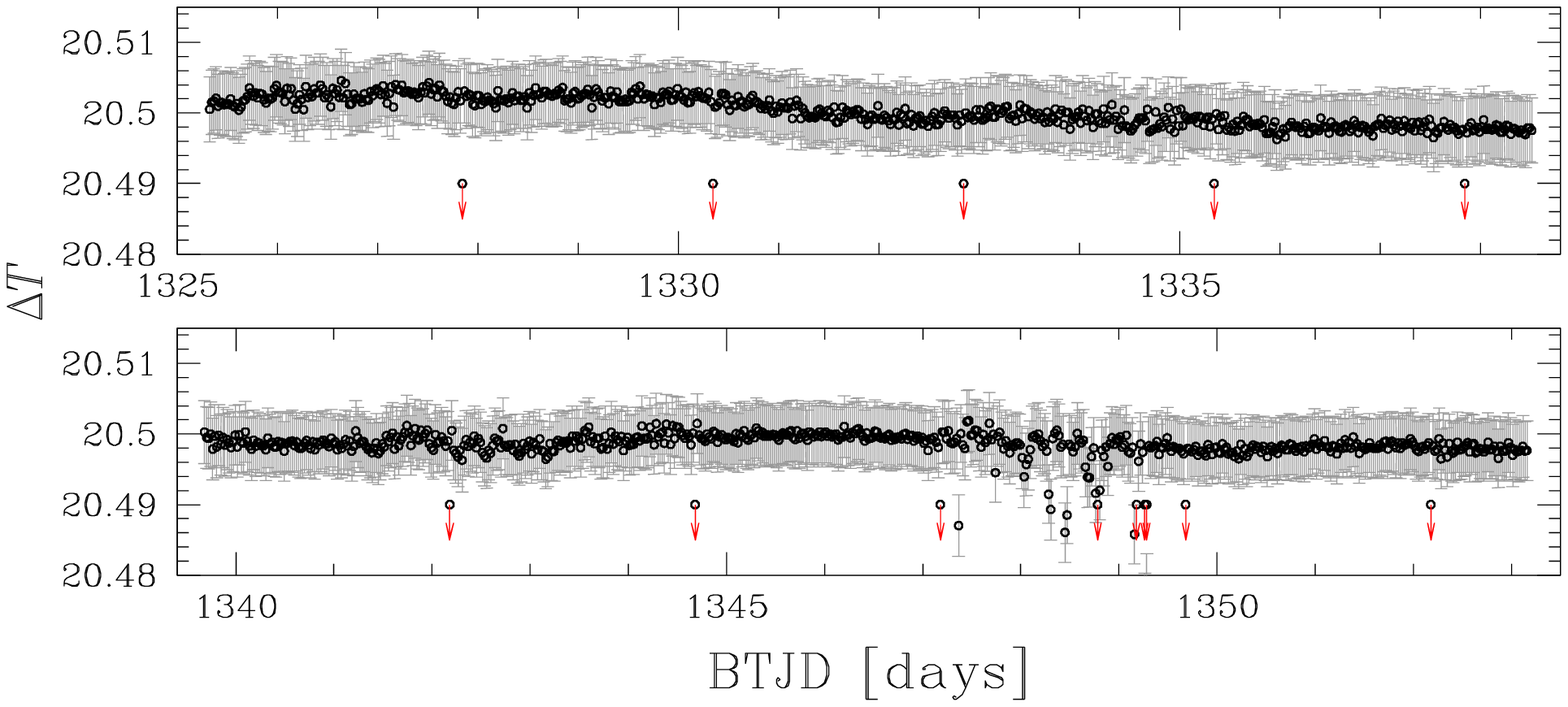}
  \caption{Variation of the zero point between \textit{TESS}
    calibrated and instrumental magnitude for Sector-1 \textit{TESS}
    observations. Red arrows indicate the points that are out of the plot limits. \label{fig3}}
\end{figure*}

\section{From Full Frame images to light curves}
\label{sec:ffi}
For this pilot work we used FFIs collected in Sector-1. FFIs are
obtained on board by co-adding series of 2\,s exposures into a 30
minute exposure time image. Observations have been carried out by
\textit{TESS} between 2018 July 25 and 2018 August 22 and cover $\sim
27.83$ days, during which 1282 usable FFIs were generated.  The target
of this work, 47 Tuc, is located on the CCD 2 of Camera 3. As
described in the next Sections, we have extracted the light curves for
the stars located in an annulus with $0.075 \le R \le 1.5$\,degree
from 47\,Tuc centre (as shown in Fig.~\ref{fig1}).

\subsection{Empirical point spread functions}

For our work it is mandatory to compute the best PSF
models for each image, both for the extraction of high precision
photometry and for the subtraction of the neighbour stars, as 
explained in Sect.~\ref{sec:lc_ext}.

The PSFs of the {\it TESS} cameras are highly variable among the large field
of view of each CCD (12$\times$12\,deg$^2$). To model the PSFs, we
used the empirical approach developed by \citet{2006A&A...454.1029A},
adapting the code used for the data collected with the ESO/MPI Wide Field Imager
to \textit{TESS} images. To take
into account the spatial variation of the PSF we divided each image
in a grid of $9 \times 9$ regions. We empirically computed the PSF
models independently in each region using bright, not saturated,
isolated stars. Each PSF model is defined on a grid of $201 \times
201$ points and is super-sampled by a factor 4 with respect to the
image pixel.

Adopting this approach, for each position of the CCD we can extract
the best, local PSF model by using a bilinear interpolation of the
four closest PSFs.

\subsection{The input catalogue}

For the extraction of the light curves we need an input catalogue that
contains the positions and the magnitudes of the stars in the analysed
field. We adopted as input catalogue the Gaia DR2 catalogue
(\citealt{2018A&A...616A...1G}). We excluded from the catalogue all
the sources with $G>17$ because they are too faint to be measured by
\textit{TESS} (the total flux is $<30$\ $e^{-}$/s). For this work, we
considered the stars located in a circular region centred in
$(\alpha_{0},\delta_{0})=(6.022329,-72.081444)$, i.e. on the centre of
47\,Tuc (\citealt{2010AJ....140.1830G}), and with radius
$R=1.5$\,degree ($\sim 2 \times$ the tidal radius of 47\,Tuc). We excluded from the input catalogue all the stars
with $R \le 0.075$\,degree because the \text{TESS} images are too
crowded and saturated within this region. In total, the input catalogue contains 16641 sources.

Figure~\ref{fig2} shows an overview of the input catalogue used to
extract and analyse the light curves: panels (a), (b), and (c) show
  the positions, the proper motions and the colour-magnitude diagram
  (CMD) of the stars analysed in this work. We plotted in red the
stars that have high probability to be 47\,Tuc members, based on
proper motions and parallaxes. We performed the 47\,Tuc member
selection as follows: first, in the vector point
  diagram (VPD) of panel (b), we selected by hand the stars that, on
the basis of their proper motion, are candidate cluster members. We
found the centre of the motion of the cluster by fitting single
Gaussians to the $\mu_\alpha \cos{\delta}$ and $\mu_\delta$
distributions of the selected stars, finding
$(\mu_{\alpha}\cos{\delta}_0,\mu_{\delta})=(5.2,-2.5)$\,mas\,yr$^{-1}$. We
selected all the stars within $3\sigma$ from the centre of the motion
and for these stars we computed the median value of the
parallax\footnote{Corrected for the offset tabulated by
  \citet{2018ApJ...861..126R}} $\pi= (0.24\pm0.06)$\,mas: we
considered candidate cluster members all the stars within
$3\sigma_{\pi}$ from the median value of $\pi$.  In the same way, we
selected Small Magellanic Cloud (SMC) members considering all the
stars within $3\sigma$ from the centre of the motion of the SMC
($(\mu_{\alpha}\cos{\delta}_0,\mu_{\delta})=(0.5,-1.2)$\,mas\,yr$^{-1}$)
and with parallaxes within $3\sigma$ from the median value $\pi= 0.03
\pm 0.07 $\,mas. The SMC high probable members are plotted in green in
panels (a), (b), and (c) of Fig.~\ref{fig2}.

We transformed the \textit{Gaia} magnitudes of the stars in the input
catalogue into \textit{TESS} magnitudes  using the relation $T-G$
versus $G_{\rm BP}-G_{\rm RP}$ illustrated in panel (d) of
Fig.~\ref{fig2}. To obtain this relation we cross-matched the
\textit{TESS} target list of the stars observed in the first seven
sectors with Gaia\,DR2 catalogue. We considered all the stars with
$G<16$ and $T<16$ and we fitted to the $T-G$ versus $B_{\rm P}-R_{\rm
  P}$ distribution a 2nd-order polynomial (magenta line in panel (d)
of Fig.~\ref{fig2}). We found:
\begin{equation*}
  T = G + a_0 + a_1\times(G_{\rm BP}-G_{\rm RP}) + a_2\times(G_{\rm BP}-G_{\rm RP})^2
\end{equation*}
where $a_0=(0.014\pm0.004)$, $a_1=(-0.643\pm 0.007)$, and $a_2=(0.055
\pm 0.003)$. Transformation from \textit{Gaia} to \textit{TESS}
magnitudes is significant  to minimise systematic residuals, during
neighbour subtraction, due to colour terms.


\begin{figure*}
  \includegraphics[width=0.995\textwidth]{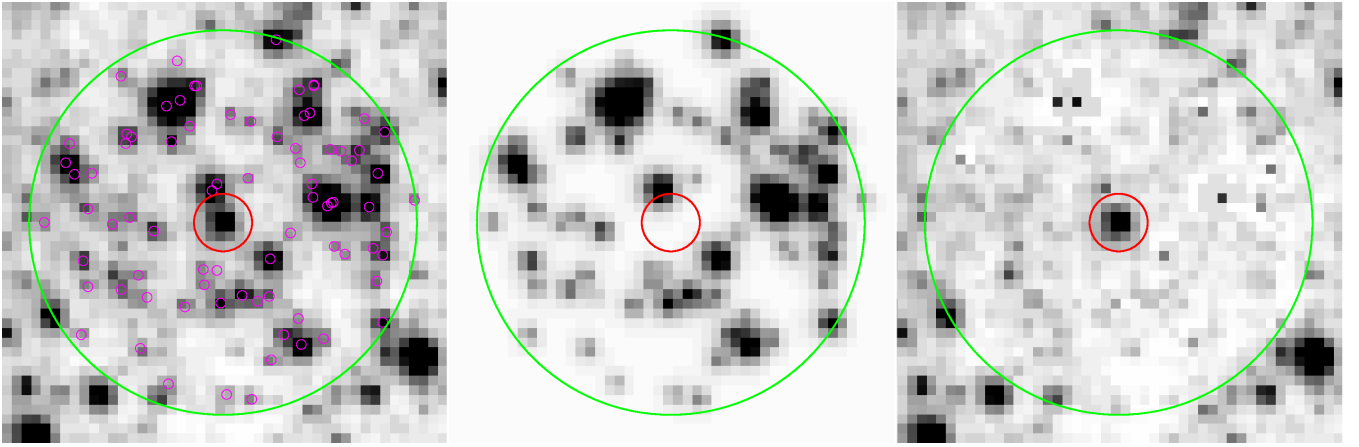}
  \caption{Procedure adopted for the subtraction of the neighbour
    stars of a target star in the input catalogue. Left-hand panel is
    one original FFI
    (\texttt{tess2018206192942-s0001-3-2-0120-s\_ffic.fits}): red
    circle marks the target star (Gaia\,DR2\,4689518492359359488), magenta
    circles are the neighbour stars in the input catalogue located
    within 20 pixels (green circle) from the target star. Middle panel
    is the image of the models of the neighbour stars, that will be
    subtracted from the original image, to obtain the
    right-hand panel. \label{fig4}}
\end{figure*}

\subsection{Light curve extraction}
\label{sec:lc_ext}

For the extraction of the light curves from \textit{TESS} images, we
used an evolved, improved version of the code \texttt{img2lc}
developed by \citet{2015MNRAS.447.3536N,2016MNRAS.455.2337N} for
ground-based images and also adopted by
\citet{2016MNRAS.456.1137L,2016MNRAS.463.1780L} and
\citet{2016MNRAS.463.1831N} for the extraction of light curves from
\textit{Kepler/K2} images of open clusters.

For each FFI, we transformed the $(\alpha,\delta)$-coordinates of the
stars in the input catalogue in the image reference system using the
WCS keywords and the distortion coefficients in the FITS header of the
images.  In particular we used the following equation to transform sky
coordinates $(\alpha,\delta)$ to geometric distortion corrected pixel
coordinates $(U,V)$ (\citealt{2002A&A...395.1061G}):
\begin{equation*}
  \binom{U}{V} = {\tt CD}^{-1}\binom{\alpha}{\delta}
\end{equation*}
where \texttt{CD} is the transformation matrix that takes into account
the rotation, scaling, and skew of the image. The \texttt{CD} elements
are listed in the FITS header.  Then, we transformed $(U,V)$ in
original pixel coordinates $(x,y)$ using two inversion distortion
polynomials, as described by \citet{2005ASPC..347..491S}:
\begin{equation*}
  x = U + \sum_{p,q}{{\tt AP}{_p}{_q} U^pV^q} \\
  y = V + \sum_{p,q}{{\tt BP}{_p}{_q} U^pV^q} 
\end{equation*}
where \texttt{AP} and \texttt{BP} are the polynomial coefficients for
the terms $U^pV^q$, and $p+q$ is the order of the polynomial. For all
the FFIs analysed in this work $p+q=4$.

For each image, we used the 200 brightest, un-saturated, and isolated stars in the input
catalogue, to derive the photometric
zero points between the calibrated \textit{TESS} magnitude $T_{\rm
  cal}$ and the instrumental magnitude $T_{\rm inst}$, $\Delta T =
T_{\rm cal}-T_{\rm inst}$. In order to compute this quantity, we
derived the instrumental magnitude of each star by fitting the local
empirical PSF, and then we computed the average value of $T_{\rm
  cal}-T^i_{\rm inst}$, with $i = 1,...,200$. Figure~\ref{fig3} shows
the variation of $\Delta T$ during the Sector-1.

For each target star in the input list and for each image, the routine
considered all the neighbour stars within a radius of 20 pixels and
transformed their calibrated magnitudes in instrumental fluxes
($e^-/s$) using the previously calculated $\Delta T$. Using the local
PSF, the transformed fluxes and the positions, the software made a
model of the neighbour stars and then subtracted it from the image.
Finally it measured the target flux using both aperture and
PSF-fitting photometry on the neighbour-subtracted image. In the case
of aperture photometry, we used 4 different aperture sizes: 1, 2, 3,
and 4 pixel radius.

Figure~\ref{fig4} shows the procedure for the subtraction of the
neighbours from a single image. Left-hand panel shows a $45\times 45$
pixel$^2$ subregion of  a FFI
 centred
on the target star.  Middle panel is the
same subregion with the models of the neighbours (located
  within a radius of 20 pixels from the target star) that are
subtracted from the original image. Right-hand panel shows the region
after the subtraction of the neighbours.

The fluxes, the epoch of the observations in TESS Barycentric Julian
Day (BTJD), the positions on the images and the local sky values were
stored in the related light curve file.

\begin{figure*}
  \includegraphics[width=0.95\textwidth,bb= 22 271 568 717]{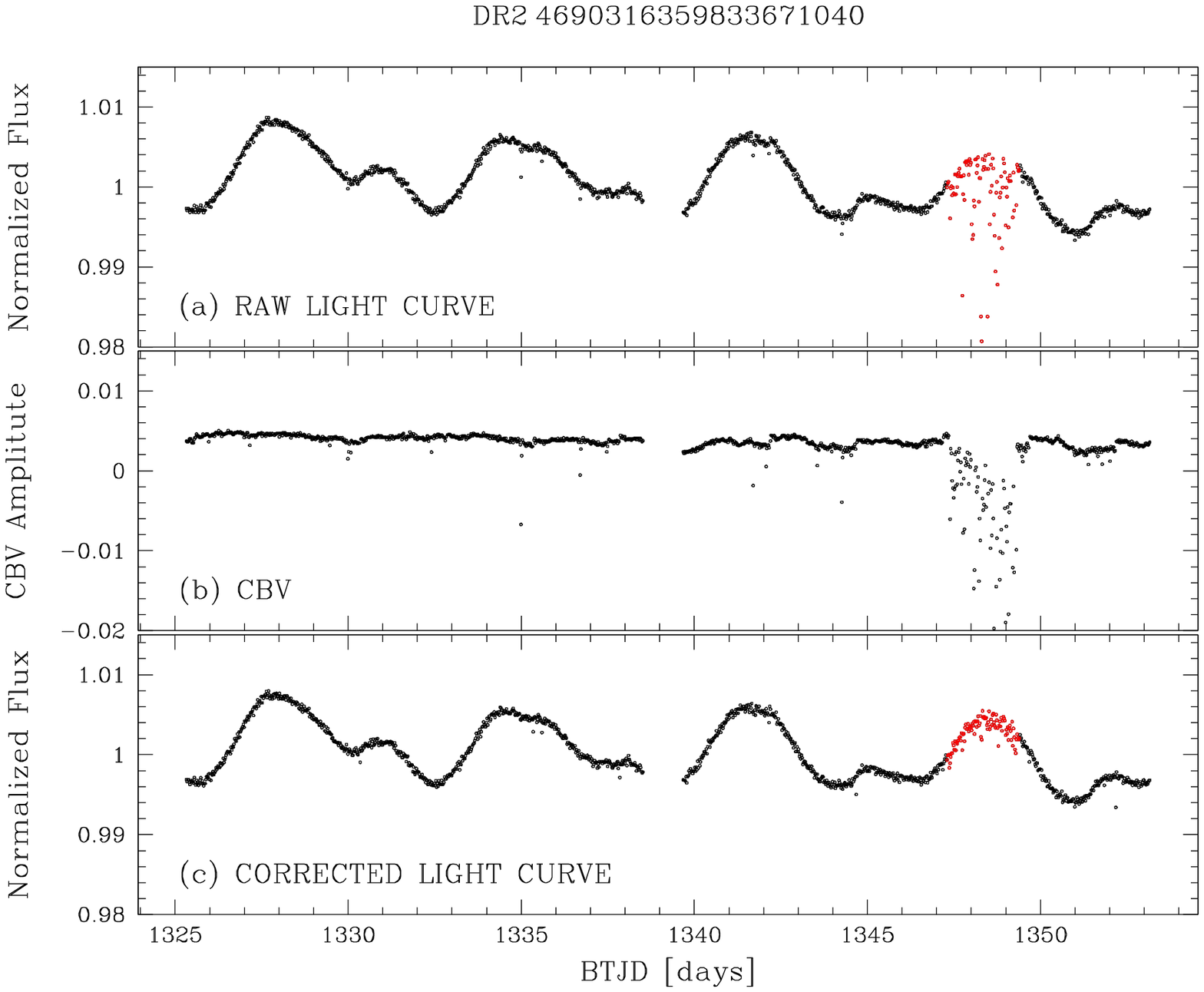}
  \caption{Procedure adopted for the correction of the light curve of
    the star Gaia\,DR2\,4690316359833671040. Panel (a) shows the normalised
    raw light curve of the star obtained with 3-pixel aperture
    photometry; panel (b) is the CBV applied to the raw light curve to
    obtain the corrected light curve shown in panel (c). Red circles
    are the epochs for which the light curve shows larger systematic
    effects, due to a pointing problem occurred during the
    Sector-1.  \label{fig5}}
\end{figure*}

\subsection{Systematic effects correction}
\label{sec:cbv}

The \textit{TESS} light curves are affected by systematic artifacts,
not correlated with the location and the luminosity of the stars on
the CCD, but associated with spacecraft, detector and
environment. Light curves of stars located on the same detector and
observed in the same sector share common systematic trends, allowing
us to model them using orthonormal functions, the so-called cotrending
basis vectors (CBVs), and correct the systematics that affect the
light curves applying to them the CBVs.

To extract the CBVs, we first computed the raw \texttt{RMS} for all
the raw light curves and for all the photometric methods.  To obtain
the raw \texttt{RMS}, we iteratively computed the median value of the light curve
and the value $\sigma$, defined as the 68.27th-percentile of the
residual from the median, clipping-out at each iteration
all the points above and below 3.5$\sigma$ from the median value;
after 10 iterations, we defined the \texttt{RMS} as the value of
$\sigma$.  For a given photometric method, we chose the magnitude
interval where the light curves have, on average, lower raw
\texttt{RMS} compared to the same light curves obtained with different
photometric methods. In this magnitude interval we selected the light
curves that showed an ${\tt RMS}<{\tt RMS}_{\rm median}+2 \sigma_{\tt
  RMS}$, where ${\tt RMS}_{\rm median}$ and $\sigma_{\tt RMS}$ are the
median of \texttt{RMS} in the considered magnitude interval, and its
standard deviation, respectively. Using these stars we extracted the
CBVs using a Principal Component Analysis. For each photometric method,
we extracted 5 CBVs, that explain 85\% of the lightcurves' variance.

To correct the light curves, we developed a routine that finds the
coefficients $A_i$ that minimise the expression:
\begin{equation}
  \label{eq:cbv}
  F^j_{\rm raw}-\sum_i{(A_i\cdot {\rm CBV_i^j})}
\end{equation}
where $F^j_{\rm raw}$ is the raw flux of the light curve at the epoch
$j=1,...,1282$, and CBV$_i$ is the $i$-th CBV, with $i=1,...,5$. To
minimise the expression (\ref{eq:cbv}), we used the
Levenberg-Marquardt method (\citealt{MINPACK-1}). We checked the final
results changing the number of CBVs applied to the light curves, and
we found that, on average, the application of just the first CBV
produced light curves with the lowest \texttt{RMS}.  In Fig.~\ref{fig5}
we show the procedure adopted for the light curve correction applied
to the star Gaia\,DR2\,4690316359833671040 ($T \sim 10$). Panel (a) shows
the 3-pixel aperture photometry: in red we highlighted the points of
the light curves that show large systematic effects due to pointing
problems occurred during the observations of the
Sector-1\footnote{\url{https://archive.stsci.edu/tess/tess_drn.html}};
panel (b) shows the CBV applied to the light curve of panel (a); panel
(c) is the cotrended light curve. It is possible to note that the
systematic artifacts due to the pointing problems are corrected.

\begin{figure*}
  \includegraphics[width=0.85\textwidth,bb= 32 325 575 701]{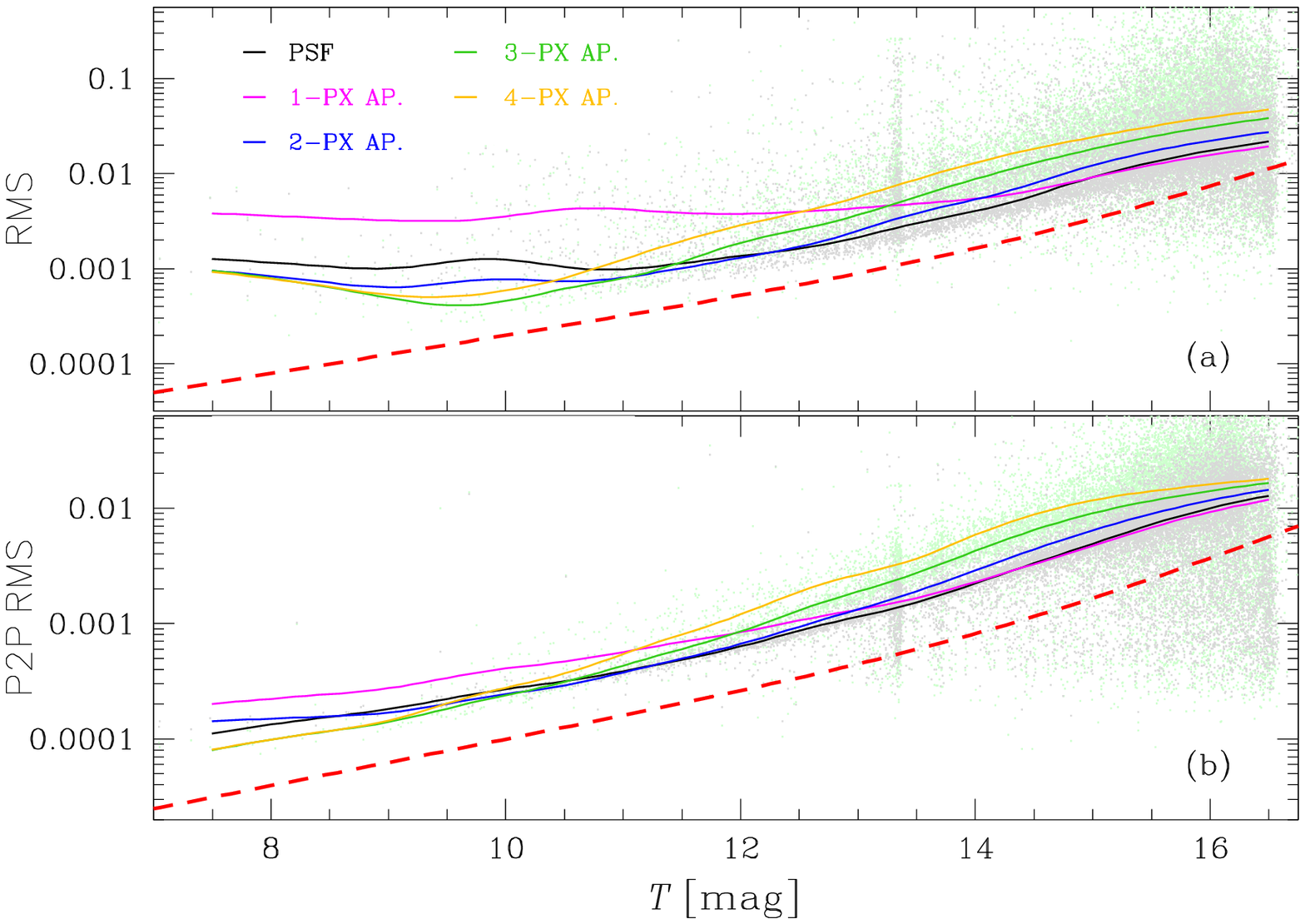} \\
  \includegraphics[width=0.85\textwidth,bb= 32 485 575 701]{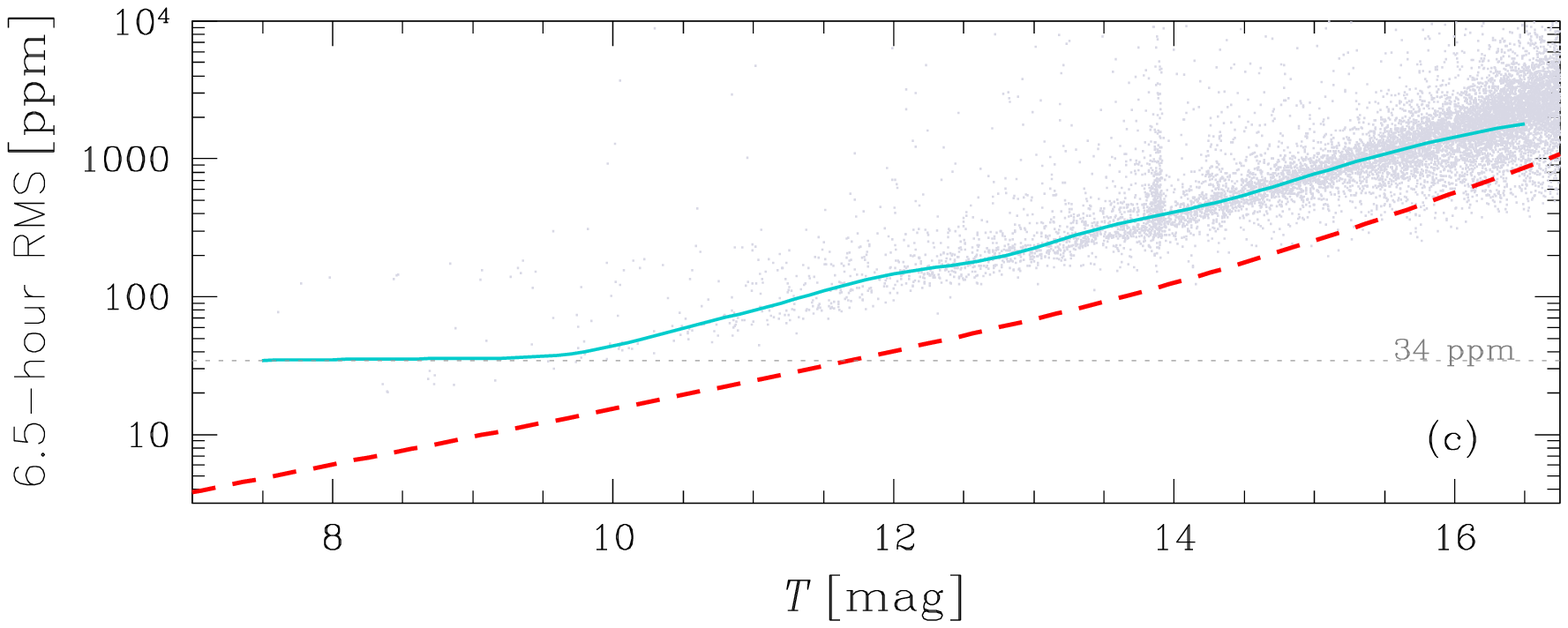} \\
  \caption{ Photometric \texttt{RMS} (panel (a)), \texttt{P2P\,RMS}
    (panel (b)), and \texttt{6.5-hour RMS} (panel (c)) as a function
    of the \textit{TESS} magnitude. In panel (a) and (b), the solid
    lines represent the average trend of the two metrics for the
    different photometric methods: PSF-fitting (black), 1-pixel
    aperture (magenta), 2-pixel aperture (blue), 3-pixel aperture
    (green), and 4-pixel aperture (orange) photometries; light green
    and grey points are the distributions for the PSF-fitting and
    3-pixel aperture photometries. Panel (c) shows the
    \texttt{6.5-hour RMS} of the stars for which the best light curve
    has been chosen: azure line represents the average trend of the
    distribution; the red dashed line is the theoretical limit
      of the three RMSs (see text for details). \label{fig6}}
\end{figure*}

\subsection{Photometric precision}
\label{sec:rms}
Panel (a) of Fig.~\ref{fig6} shows the \texttt{RMS} as a function of
the \textit{TESS} magnitude for all the stars for which we extracted
the light curves. The \texttt{RMS} is computed using the cotrended
light curves and adopting the procedure described in
Sect.~\ref{sec:cbv}. For each photometric method, we divided the
\texttt{RMS} distribution in bins of 0.75 $T$ magnitude, and, within
each bin, we calculated the 3.5$\sigma$-clipped average (with 10
iterations) of the \texttt{RMS}. We interpolated the mean \texttt{RMS}
values with a cubic spline. We show in panel (a) of Fig.~\ref{fig6}
the interpolated values with black, magenta, blue, green, and orange
lines for PSF-fitting, 1-pixel, 2-pixel, 3-pixel, 4-pixel aperture
photometries, respectively. The dashed red line is the
  theoretical \texttt{RMS} limit, obtained considering all the sources
  of noise (shot noise, sky, Readout Noise, and dark current), and
  adopting an average Readout Noise of 9\,$e^-$/pixel, a dark
  current of 1\,$e^-$/s/pixel\footnote{As reported in
    the
    \href{https://archive.stsci.edu/files/live/sites/mast/files/home/missions-and-data/active-missions/tess/_documents/TESS_Instrument_Handbook_v0.1.pdf}{
      TESS Instrument Handbook}.}, an average sky value of
  100\,$e^-$\,s$^{-1}$, and an average aperture of 2 pixels. In the
bright regime ($T\lesssim10.5$) 3-pixel and 4-pixel aperture
photometries give the lowest \texttt{RMS}; for stars with $10.5
\lesssim T < 13.5$ 2-pixel aperture photometry is the best photometry;
in the faint regime 1-pixel and PSF-fitting photometry give the best
results.  For completeness, we plotted in grey and light green the
\texttt{RMS} distributions for PSF-fitting and 3-pixel aperture
photometries, respectively.

Because the simple \texttt{RMS} parameter is affected by the
variability of the stars, we computed the point-to-point \texttt{RMS}
(\texttt{P2P\,RMS}), defined as the 68.27-th percentile of the
distribution of the residuals from the median value of $\delta F$,
where $\delta F_j= F_j-F_{j+1}$, and $F_j$ and $F_{j+1}$ are the flux
values at the epoch $j$ and $j+1$, and $j=1,...,1281$. We derived the
mean trends of the \texttt{P2P\,RMS} distributions for the different
photometric methods as done for the \texttt{RMS}. For the following
analysis, we used these mean \texttt{P2P\,RMS} trends to identify, for
each star, the best photometry: for each star in the input catalogue
with a given $T$ magnitude, we selected the light curve obtained with
the photometric method that at the given $T$ returns the lower mean
\texttt{P2P\,RMS}.  In this phase we also excluded all the light
curves that have $<300$ photometric points and whose mean magnitude
differs from that expected from the input catalogue $>2$\,magnitudes.

Panel (c) of Fig.~\ref{fig6} shows the \texttt{6.5-hour RMS}
distribution (grey points) for the previously selected light
curves. This transit noise \texttt{6.5-hour RMS} is obtained as
described by \citet{2011ApJS..197....6G,2015AJ....150..133G} and
\citet{2016PASP..128g5002V}: we flattened each light curve
interpolating to it a 5th-order spline with 65 break points and
removing out the outliers. We computed the standard deviation of a
running mean with a window length equal to 6.5-hour (13 points). The
azure line in panel (c) is the mean trend of the \texttt{6.5-hour RMS}
distribution, calculated as previously described: the trend reaches a
minimum at $\sim 34$\,ppm. For $T<10$ the distribution is bi-modal:
stars that are not variable or whose period variability is
$>6.5$\,hours have \texttt{6.5-hour RMS}$\sim 20$\,ppm. The stars with
\texttt{6.5-hour RMS} $\gtrsim$ 30 are highly variable on short
timescales and the flattening is not perfect. For stars with $10
\lesssim T \lesssim 14$ we are able to detect exoplanets transits with
depth $< 1$\,mmag, while for $T \gtrsim 14$ it is possible to find
transits with depth $< 0.01$\,mag.

\section{Candidate exoplanet transit finding}
\label{sec:candexo}
We searched for transiting signals among the light curves selected in
the previous section. As described in Sect.~\ref{sec:rms}, we
flattened all the light curves using 5-th order splines defined on 65
knots. For each light curve, we removed out the outliers and the
photometric points with $1347.4<t_{\rm BTJD}<1349.4$, whose
photometry is affected by pointing problems and that, despite the
cotrending,  still have some residual systematic artifacts.

For each light curve we extracted both the Box-fitting Least-Squares
(BLS, \citealt{2002A&A...391..369K}) and the Transit-fitting Least
Squares (TLS, \citealt{2019A&A...623A..39H}) periodograms. For both
techniques we searched for transiting objects having period $0.6\leq
P \leq 14$\,days, and for each star we extracted the parameters
useful to discriminate between light curves with and without candidate
transits, such as the Signal Detection Efficiency (SDE), the
Signal-to-Noise Ratio (SNR), the depth of the transit.  To detect
candidate transits we performed two independent analysis on BLS and
TLS outputs and at the end we joined the results.

\begin{figure}
  \includegraphics[width=0.53\textwidth,bb= 44 144 505 714]{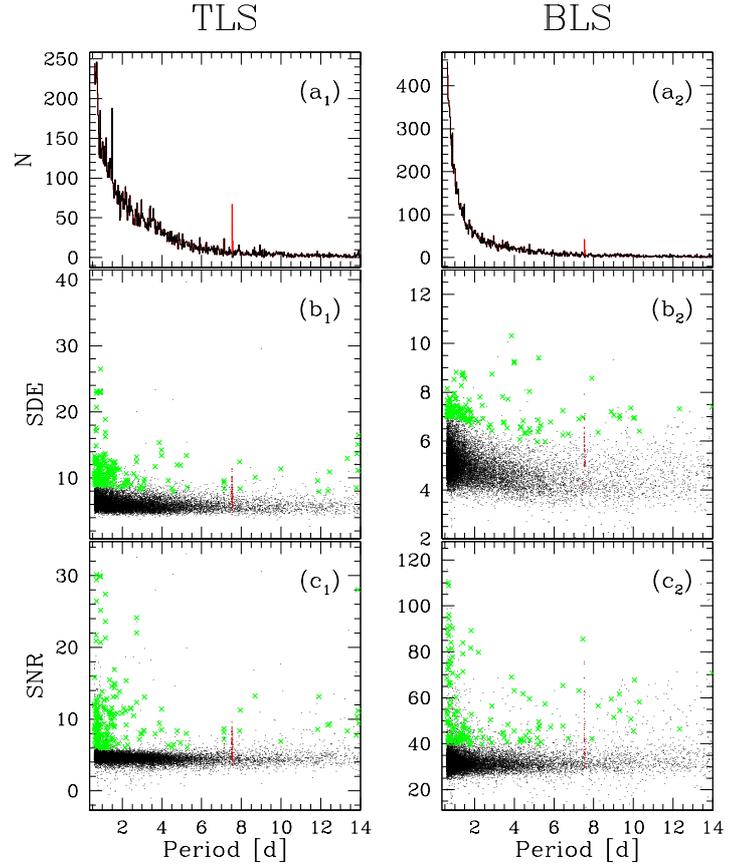}
  \caption{Overview on the selection of candidate transiting
    exoplanets adopted using TLS (left panels) and BLS (right panels)
    outputs. Panels (a) show the histogram of the best periods found
    using the different routines. Panels (b) are the SDE distributions
    as a function of the period, while panels (c) show the SNR
    distributions as function of the period. The red spike in panels
    (a) and the red points in panel (b) and (c) are the stars excluded
    in the spike suppression procedure. Green points are the stars
    that passed the selection criteria. \label{fig7}}
\end{figure}

In Fig.~\ref{fig7} we summarise the procedure adopted for the
selection of candidate transiting exoplanets both for TLS (left
panels) and BLS (right panels) technique. As described in
\citet{2015MNRAS.447.3536N}, we excluded spurious periods due to
systematic errors by constructing the histogram of the period and
removing the stars that form the spikes in the histogram: the red
spike ($\sim 7.5$\,d) in panels (a$_1$) and (a$_2$) represents the
stars we removed. In a second step, we divided the distributions of
SDE (SNR) in period intervals $\delta P=0.5$\,d and, within each bin,
we calculated the 3$\sigma$-clipped mean value of SDE (SNR),
$\overline{{\rm SDE}}$ ($\overline{{\rm SNR}}$), and its standard
deviation $\sigma_{\rm SDE}$ ($\sigma_{\rm SNR}$). We interpolated the
mean points with a spline and we saved all the stars with ${\rm SDE}>
\overline{{\rm SDE}}+3 \times \sigma_{\rm SDE}$ and ${\rm SNR}>
\overline{{\rm SNR}}+3 \times \sigma_{\rm SNR}$. We excluded all the
stars having transit depths $>5$\%. In panels (b) and (c) the stars
that passed the selection criteria are plotted in green: we recovered
186 stars and 145 stars in the case of TLS and BLS, respectively, with
63 stars in common.

Finally we visually inspected the candidate transits that passed the
selections to exclude false alarms. We found 7 interesting objects on
which we performed a series of tests to confirm or exclude the planetary nature.

\begin{figure*}
  \includegraphics[width=0.45\textwidth,bb= 18 148 333 714]{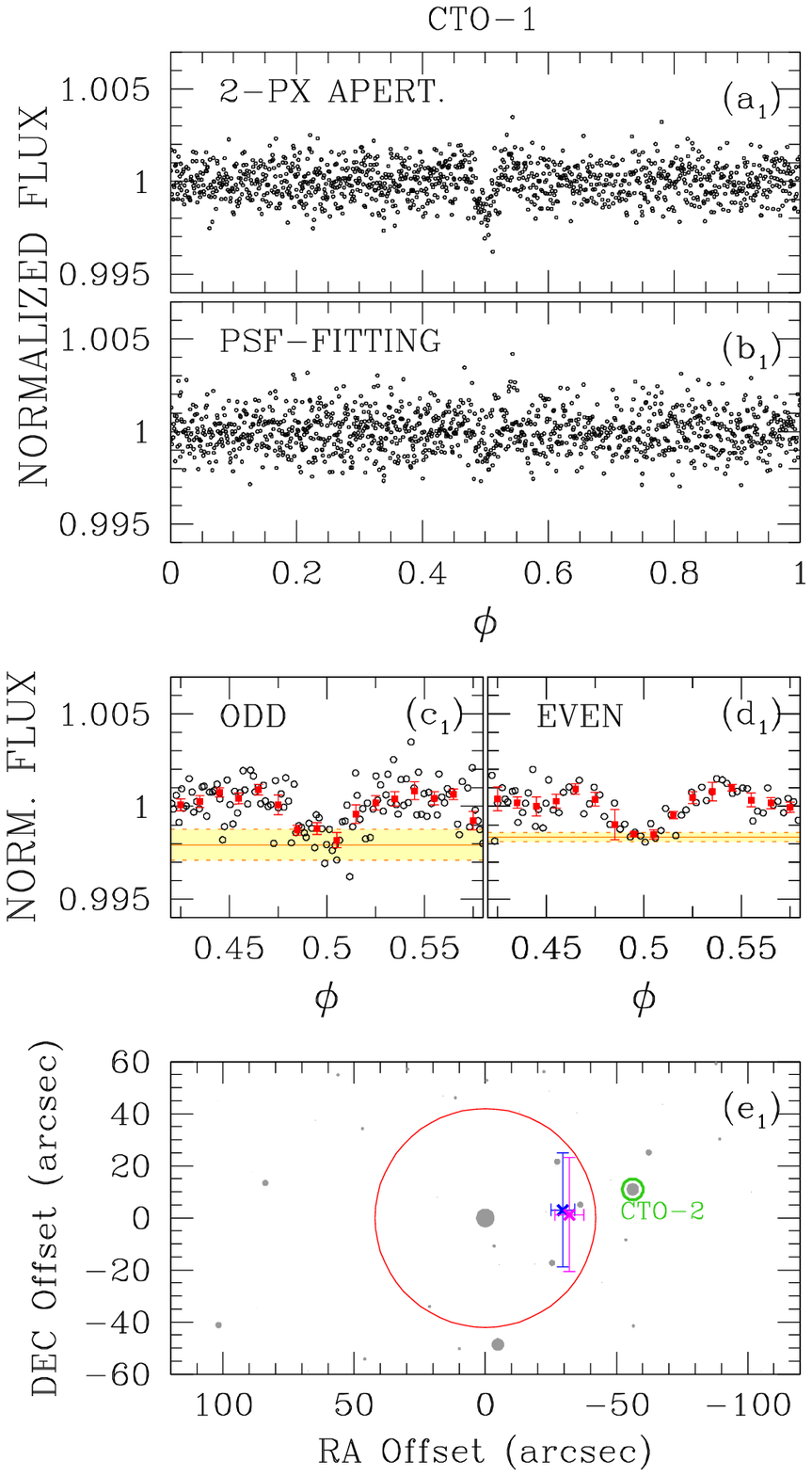}
  \includegraphics[width=0.45\textwidth,bb= 18 148 333 714]{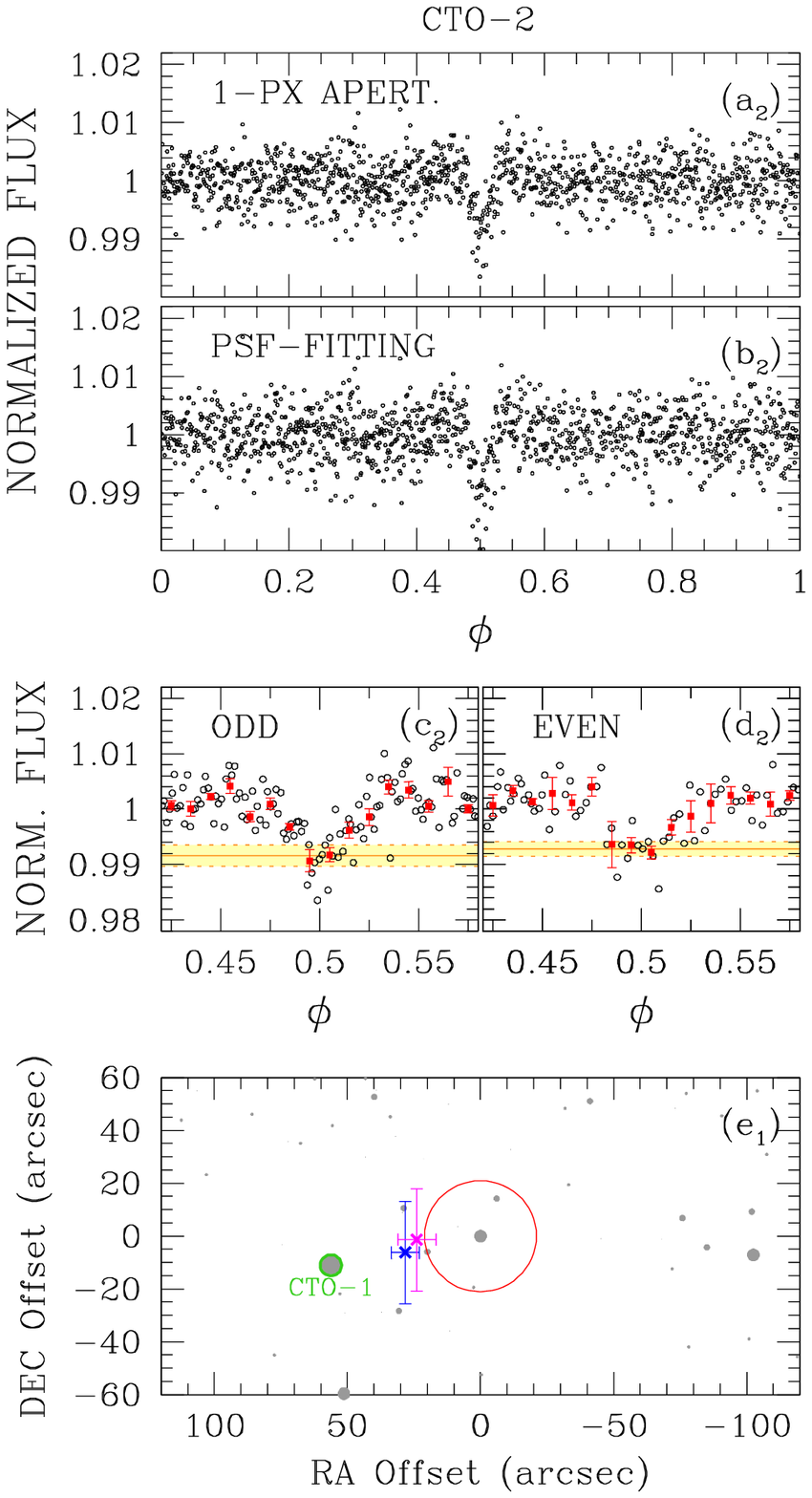} 
  \caption{Procedure adopted for the vetting of the candidate
    exoplanets CTO-1 and CTO-2. Panels (a) show the
    best-photometry phased light curves for the two targets. Panels
    (b) are the phased light curves obtained with PSF-fitting
    photometry. Panels (c) and (d) plot the phased light curves for odd
    and even transits, respectively: orange horizontal lines represent
    the depth of the transits, in yellow the 1$\sigma$ zone.  Panels
    (e) show the centroid offsets calculated using the centroids from
    the out-of-transit image (magenta crosses) and the position from
    Gaia\,DR2 (blue crosses); the red circle is the photometric
    aperture adopted, grey circles are all the stars in the Gaia DR2
    catalogue.
    \label{fig8}}
\end{figure*}

\begin{figure}
  \includegraphics[width=0.45\textwidth,bb= 18 148 333
    714]{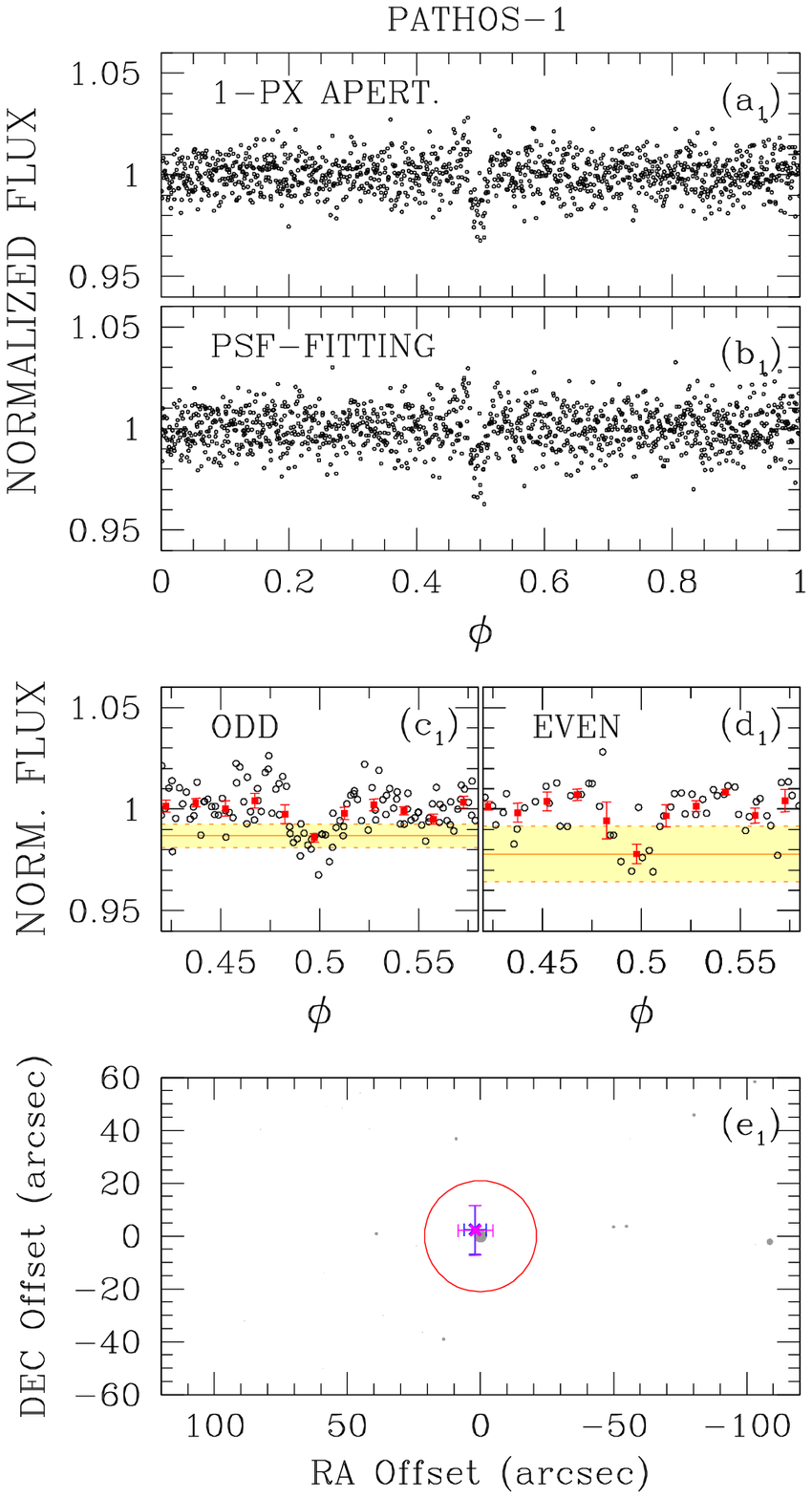}
  \caption{As in Fig.~\ref{fig8}, but for PATHOS-1    \label{fig9}}
\end{figure}

\subsection{Neighbour contamination}
As a first test, we checked if the candidates are contaminated by
close stars. We considered all the stars within the input
catalogue that are at a distance $<100$\,arcsec ($\lesssim 5$\,pixels)
from the target candidate and we checked their periods and phased
light curves. In this way we excluded 4 stars that are blended with
close eclipsing binaries.

Among the 3 stars that succeed the check, two of them are close stars
showing the same (interesting) signal. We will analyse all of
them in the next section.

\subsection{Vetting and modelling of the candidates }

In this section we verified that the selected stars are well-behaved candidate
transiting exoplanets. In order to do this, we did the following
tests: (i) we checked the position of the centroid of the star during
and outside the transit events; (ii) we compared the depth of the odd
and even transits; (iii) we compared the best photometry with that
obtained using PSF-fitting, less affected by the contamination
effects. The following sections will be dedicated to the vetting of
the candidate transiting objects (CTOs).

\subsubsection{Two close candidates}

The first two candidates are two close stars at a distance $\delta r
\simeq 57.2$\,arcsec, and show similar signals in their light
curves. The brighter star (Gaia\,DR2\,4689801887175745536, hereafter CTO-1)
has $T \sim 12.2$ and, as explained in Sect.  \ref{sec:rms}, the best
light curve is obtained using the 2-pixel aperture photometry. The
fainter star (Gaia\,DR2\,4689813608144062080, hereafter CTO-2) has $T \sim
14.3$, and 1-pixel aperture photometry gives the best
result. The two stars present a periodic transit event with period $P
\sim 3.982$\,d, but different depth: for CTO-1, the transit depth is
$\sim 0.2$\,\% (panel (a$_1$) of Fig.~\ref{fig8}), while for CTO-2 the
transit is $\sim 4$ times deeper, as shown in panel (a$_2$) of
Fig.~\ref{fig8}.

To verify if the signal belongs to CTO-1 or CTO-2, for both the
targets we compared the best-aperture photometry with the PSF-fitting
photometry (panels (b) of Fig.~\ref{fig8}), that is less affected by
contamination problems. As shown in panel (b$_1$) of Fig.~\ref{fig8},
even if the scatter of the PSF-fitting light curve of CTO-1 is
similar to that of 2-pixel aperture photometry ($\sigma({\rm PSF})\sim
1060$\,ppm versus $\sigma({\rm APER})\sim 970$\,ppm), the PSF-fitting
photometry shows a less evident transit, with depth $\sim 0.1$\%. On
the other hand, comparing the PSF-fitting (panel (b$_2$)) and 1-pixel
aperture (panel (a$_2$)) photometries of CTO-2, it is possible to
confirm the presence of the transits in both light curves, with the
same depth.

As shown in Fig.~\ref{fig8}, we compared the depth of odd (panels (c))
and even (panels (d)) transits, to exclude the hypothesis that the
transits events are eclipses of an eclipsing binary. We calculated the
mean depth of odd/even transits (orange continuous lines in panels
(c)) and its standard deviation (yellow strips): for CTO-1, we
found that the odd and even transits have a depth of $2.1 \pm 0.8$
mmag and $1.6 \pm 0.3$ mmag, respectively; CTO-2 have odd and even
transits depth of $8.3 \pm 1.9$ mmag and $7.2 \pm 1.3$ mmag. For both
 stars, the depth of odd and even transits overlaps within 1$\times
\sigma$.

We analysed the centroid offset using the technique described by
\citet{2018PASP..130f4502T}. Briefly, we selected the images
corresponding to the in-of-transit events and the images near the
transit events (out-of-transit). For each transit, we calculated the
mean in- and out-of-transit images, and then we stacked together all
the mean in- and out-of-transit images. We calculated the difference
image subtracting pixel-by-pixel the in-of-transit stack to the
out-of-transit stack. We calculated the photocentres on the
out-of-transit and on the difference images.  We calculated two
offsets: (1) the difference between the centroid obtained from the
out-of-transit and the difference images, and (2) the difference
between the Gaia\,DR2 position and the centroid from the difference
image. The offset locates the source of the transit signature.

\begin{table}
  \caption{Transit parameters for the candidate exoplanet.}
    \label{tab1}
    \resizebox{0.449\textwidth}{!}{
      { \renewcommand{\arraystretch}{1.2}
\begin{tabular}{l c}
  \hline
  \multicolumn{2}{c}{PATHOS-1} \\
  \hline
      {Gaia DR\,2 ID} & 4702085154340085248 \\
      
      {$\alpha_{\rm J2000}$ ($^\circ$)} & 2.87965146  \\
      {$\delta_{\rm J2000}$ ($^\circ$)} & $-$70.98693522 \\
      {$T$ (mag)} & 15.3 \\
      {$R_\star$ ($R_\odot$)} & $0.86 \pm 0.02$ \\
      {$M_\star$ ($M_\odot$)} & $0.78 \pm 0.02$ \\
      {$\rho_\star\, (\rho_\odot)$} & $1.24_{-0.06}^{+0.07}$\\
      {Period (d)} & $3.8582_{-0.0014}^{+0.0017}$  \\
      {$T_0$ (BTJD) } & $1339.677 \pm 0.003$ \\
      {$R_{\rm p}/R_\star$} & $ 0.152_{-0.009}^{+0.011}$ \\
      {$R_{\rm p}$ ($R_{\rm Jup}$)} & $1.27_{-0.07}^{+0.10}$  \\
      {$a/R_\star$} & $11.11_{-0.22}^{+0.17}$\\
      {$i$ ($^\circ$)} & $89.2_{-0.5}^{+0.8}$ \\
      {$T_{14} $~(min)} &  $181_{-4}^{+5}$ \\
      {$u1  $} &  $ 0.42\pm 0.10$ \\
      {$u2  $} &  $ 0.18\pm 0.10$ \\
  \hline
\end{tabular}
}

      }

\end{table}

Panels (e) of Fig.~\ref{fig8} shows the out-of-transit centroid
analysis for the two stars. In each panel, the target star is centred
in (0,0) and the red circle is the photometric aperture used to extract the light curve; in grey
are all the sources in the Gaia\,DR2 catalogue, and magenta and blue
crosses are the centroid offsets calculated using the centroids from
the out-of-transit image and the position from Gaia\,DR2,
respectively. According to the centroids, the transit events do not
occurs on CTO-1 or CTO-2, but to a very faint star ($G >17.2$)
between the two candidates.

We searched in literature whether already discovered eclipsing binaries with magnitude $G>17$ are located  in that sky regions:
\citet{2016AcA....66..421P}, in a study of eclipsing binaries in the
Magellanic Clouds using OGLE data (\citealt{1992AcA....42..253U}),
found an eclipsing binary (OGLE-SMC-ECL-6193) with period $P \sim
7.9733$, so twice the period of the two candidates, $G \sim 17.25$ and
located exactly where the centroids are shown in panels (e) of
Fig.~\ref{fig8}.  The depth of the primary eclipses is $\sim 0.3$
magnitudes, while that of the secondary eclipses is $\sim 0.2$ . So,
given the differences in magnitudes between the two candidates and the
eclipsing binary and the fraction of light of the eclipsing binary
that falls within the aperture of the two candidates, we expect an
induced transit of depth $\sim 0.1$\% and $\sim 1$\% for CTO-1 and
CTO-2, respectively.

We conclude that the two candidates are false positive generated by the
blending with a fainter eclipsing binary located between them.

\begin{figure}
  \includegraphics[width=0.55\textwidth,bb= 20 151 430 707]{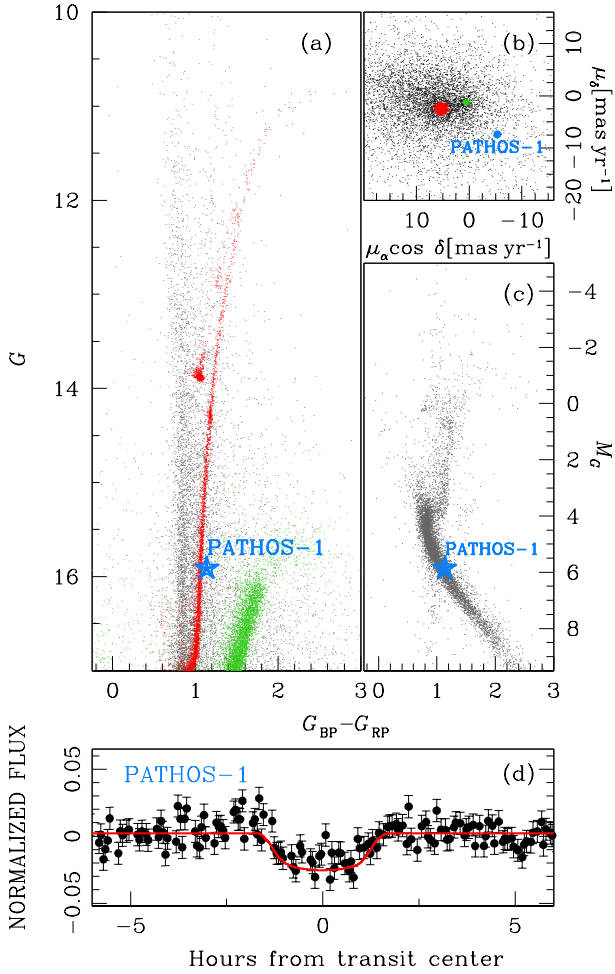}
  \caption{ Panel (a) shows the $G$ versus $G_{\rm BP}-G_{\rm RP}$
    CMD: in red are the stars that have high probability to be 47\,Tuc
    members, in green the stars with high probability to belong to the
    SMC, in grey the field stars of the Milky Way.  Panel (b) is the
    VPD for 47\,Tuc, SMC, and Milky Way stars plotted in red, green
    and grey, respectively. Panel (c) shows the absolute
    magnitude $M_{G}$ versus $G_{\rm BP}-G_{\rm RP}$ CMD for the Milky
    Way stars. In all the panels, the position
    of the candidate exoplanet PATHOS-1 is shown in azure.  Panel (d) is the phased
    light curve of PATHOS-1; in red the best-fit model (see
    text for details). \label{fig10}}
\end{figure}

\subsubsection{A candidate exoplanet: PATHOS-1}

Figure~\ref{fig9} shows the vetting procedure for the candidate
transiting exoplanet PATHOS-1 ($T \sim 15.3 $, $P \sim 3.86$\,d). The
comparison between the best photometry (1-pixel aperture, panel
(a$_1$)) and the PSF-fitting photometry shows that it is possible to
identify the transit events in both of the light curves.  Comparing
the odd ($1.3\pm0.6$\,\%) and even ($2.2 \pm 1.3$\,\%) transit events
(panels (c$_1$) and (d$_1$)) we found that the depths are compatible
within 1$\times \sigma$.  From the analysis of out-of-transit
centroids (panel (e$_1$) of Fig.~\ref{fig9}), we confirm that the
transit events occur on PATHOS-1.

PATHOS-1 is a field star located on the main sequence (MS) of the Galaxy (see
panels (a), (b), and (c) of Fig.~\ref{fig10}).


We extracted transit parameters using a modified version of the code
used by \citet{2019MNRAS.484.3233B} and
\citet{2019arXiv190401591B}. It selects a portion of the light-curve
around each transit of about $\pm 2 \times T_{14}$ ($T_{14}$ is the
duration of the transit from the TLS) from the linear ephemeris
determined with TLS.  This code models the transit with the
batman-package (\citealt{2015PASP..127.1161K}) and computes the
posterior distribution with \texttt{emcee} (affine invariant
Markov-Chain Monte Carlo sampler,
\citealt{2013PASP..125..306F}).

The code fits as common parameters the stellar density $\rho_\star$ (in
Solar unit), the base-2 logarithm of the period of the planet ($\log_2
P$), the radii ratio ($k=R_\mathrm{p}/R_\star$), the impact parameter
($b$), the quadratic limb darkening (LD) parameters, $q_1$ and $q_2$,
as proposed by \citet{2013MNRAS.435.2152K}, a base-2 logarithm of a
jitter term ($\log_2 \sigma_j$), the reference transit times ($T_0$).
For each transit it fits a detrending polynomial of third order
(coefficients $c_0$, $c_1$, $c_2$) and the time of the centre of the
transit ($TT$).  It this way it computes simultaneously each transit
time ($TT$) and the linear ephemeris (given by the period and $T_0$).
We fixed the eccentricity $e$ to 0 and the argument of the pericenter
$\omega$ to 90 deg.  We used uniform-uninformative priors within
conservative boundaries for all the parameters, but Gaussian priors
for the stellar density and LD parameters.  The prior for the density
has been computed from the stellar radius ($R_\star$) and mass ($M_\star$); from
the Gaia\,DR2 catalogue the star has $R_\star=0.86\pm0.02\, R_\odot$
and, using mass-luminosity relation $L \propto M^{3.5}$, we found
$M_\star=0.78\pm0.02\, M_\odot$.  The quadratic LD parameters have
been determined from the \citet{2018A&A...618A..20C} table for
$T_{\rm eff}=4985\pm59$~K (from Gaia\,DR2) and we adopted priors with
a Gaussina conservative error of about 0.1 for both the parameters.

We run \texttt{emcee} with 72 walkers and for 20\,000 steps.  We removed the
first 2000 steps as burn-in (checked visually the convergence of the
chains) and we used a pessimistic thinning factor of 100.  From the
posterior distribution we computed the physical posteriors and
computed the uncertainties as the high density interval (HDI) at
$68\%$ ($1\sigma$ equivalent).  We computed the median of the physical
posterior distribution as the best-fit transit parameters (reduced
$\chi^2 \sim 0.92$).  We found that the period of the candidate
exoplanet is $P \sim 3.8582$~d and $k \sim 0.152$, i.e., $R_\mathrm{p} \sim 1.27\,
R_\mathrm{Jup}$. Table~\ref{tab1} gives all the physical
parameters with errors.

Panel (d) of Fig.~\ref{fig10} shows the phased
light curve and, over-imposed in red, the best fitting model.

\begin{figure*}
  \includegraphics[width=0.995\textwidth,bb= 19 220 588 700]{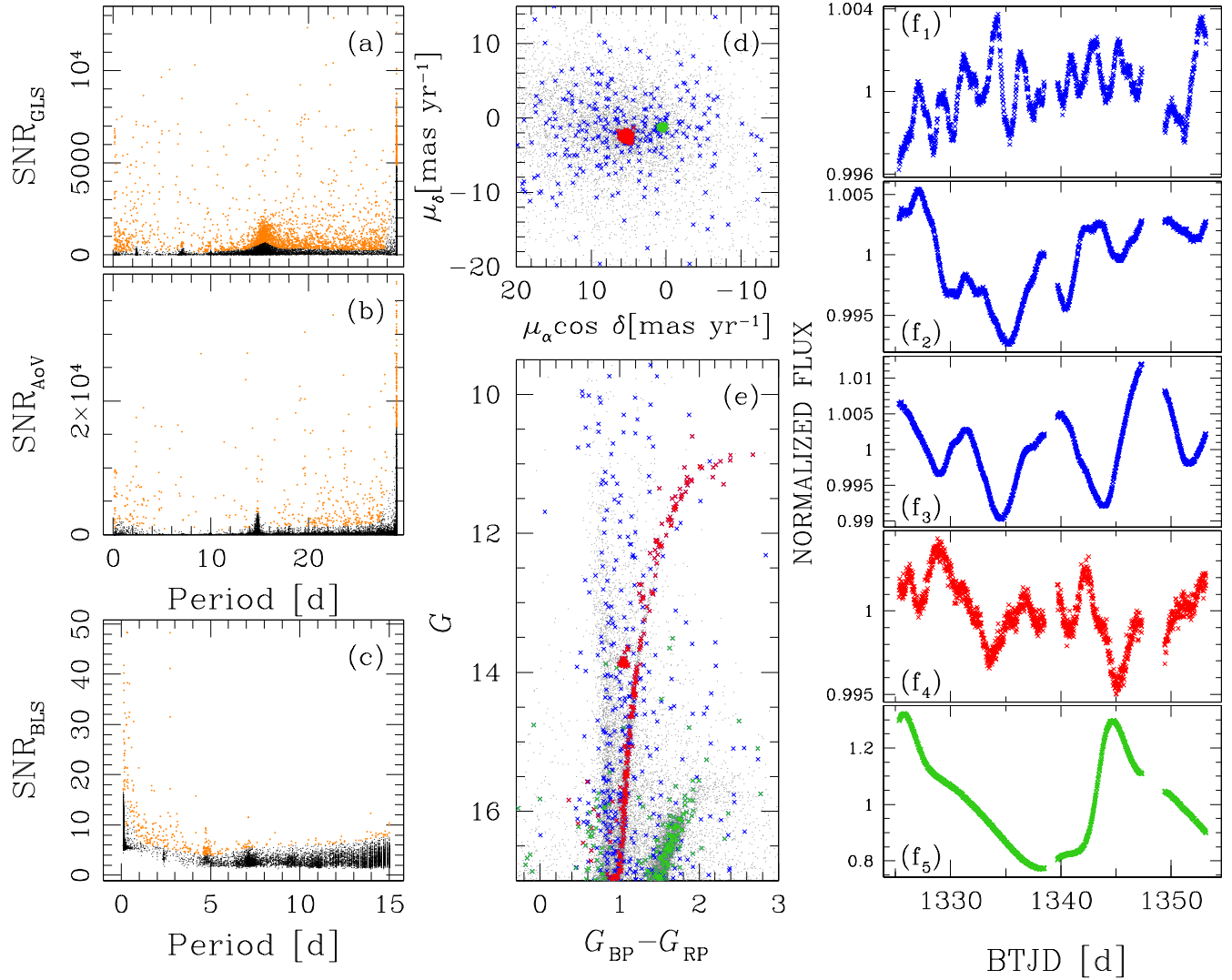}
  \caption{Procedure adopted for the selection of candidate variable
    stars. Panels  (a), (b), and (c) show
    the selection in the SNR versus Period plane for the GLS,
      AoV, and BLS periodograms, respectively: in orange are the
    stars that have high probability to be variables. Panels  (d) and (e) are the VPD and the $G$ versus
    $G_{\rm BP}-G_{\rm RP}$ CMD, respectively: in blue, red, and green
    are the candidate variable stars that belong to field, 47\,Tuc,
    and SMC, respectively. Panels  (f) are some
    examples of light curves of stars with high probability to be
    variables: the light curves are colour-coded as in panel (d) and (e).  \label{fig11}}
\end{figure*}

\begin{figure*}
  \includegraphics[width=0.995\textwidth, bb=15 34 579 576]{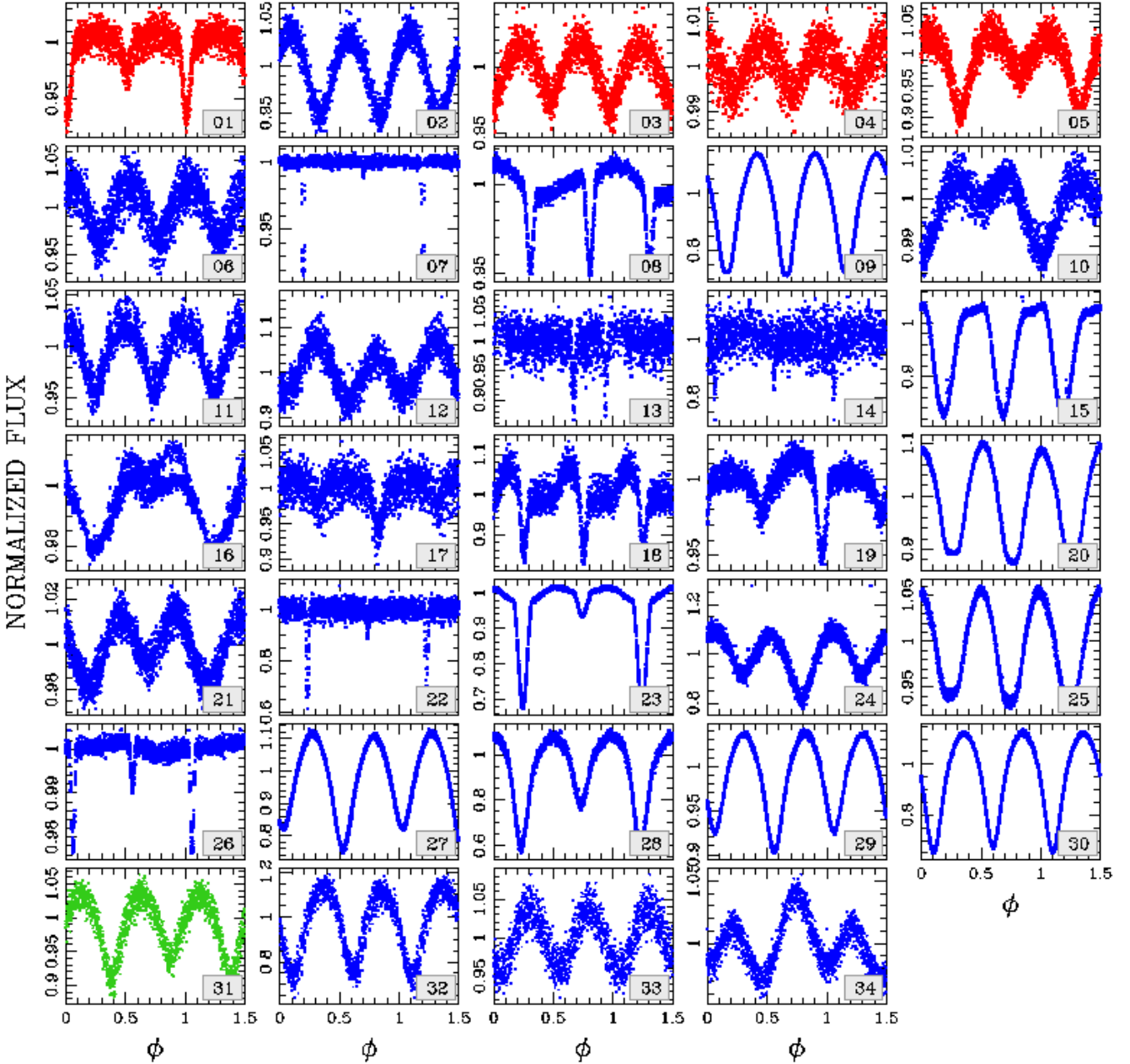}
  \caption{Light curves of eclipsing binaries found in the field analysed in this work. Light curves
    are colour-coded as in Fig.~\ref{fig11}.  \label{fig12}}
\end{figure*}

\begin{table*}
  \caption{List of eclipsing binaries}
    \label{tab2}
    \resizebox{\textwidth}{!}{
      \begin{tabular}{l c c c c c c c c c c c c}
  \hline
  {\bf \# EB} & {\bf $\alpha$(J2000)} & {\bf $\delta$(J2000)} & {\bf Gaia DR\,2 ID} & {\bf $T$} & {\bf $G$} & {\bf $G_{\rm BP}$} & {\bf $G_{\rm RP}$} & {\bf $\mu_{\alpha}\cos \delta$} & {\bf $\mu_\delta$} & {\bf P} & {\bf Note} & {\bf Ref.}\\
  {} & {($^\circ$)} & {($^\circ$)} & {} & {} & {} & {} & {} & {(mas\,yr$^{-1}$)} & {(mas\,yr$^{-1}$)} & {(d)} & &  \\
  \hline
%
01   &     6.537000594   &   -72.11708689  & 4689629611756844672        & 15.65        & 15.84        & 15.95        & 15.59 &        5.267 &       -1.957 &      1.150726 &	V46, 0063	&      (1,2)                        \\ 
02   &     5.697991131   &   -72.22145320  & 4689625041910756480        & 15.11        & 15.53        & 15.84        & 15.04 &       11.575 &       -1.031 &      0.278849 &	V50		&      (1)                          \\ 
03   &     6.543979192   &   -72.18550401  & 4689628203007617792        & 16.17        & 16.46        & 16.52        & 15.98 &        5.660 &       -2.563 &      0.378792 &	V45, 6257	&      (1,2)		             \\ 
04   &     5.825314808   &   -72.31158497  & 4689619853590073216        & 13.33        & 13.88        & 14.34        & 13.26 &        5.802 &       -2.943 &      0.383336 &	V49		&      (1)                          \\ 
05   &     6.679190520   &   -72.25576887  & 4689579961931616640        & 16.38        & 16.72        & 16.93        & 16.29 &        5.911 &       -2.183 &      0.446193 &	V53, 6261	&      (1,2)		             \\ 
06   &     6.749464138   &   -71.91938968  & 4689646757262208000        & 15.43        & 15.88        & 16.01        & 15.16 &        7.993 &        2.603 &      0.347537 &	KalE1  		&      (1)                          \\ 
07   &     5.413085792   &   -72.52089953  & 4689547010942141056        & 13.47        & 14.00        & 14.42        & 13.41 &        9.254 &      -10.723 &      5.227904 &	                &                                   \\ 
08   &     6.355276472   &   -72.54985787  & 4689552783378799744        & 13.75        & 14.41        & 14.96        & 13.66 &       12.478 &        6.540 &      0.677519 &	0057      	&      (2)                          \\ 
09   &     4.340046990   &   -71.91583531  & 4689821167286415744        &  9.74        & 10.02        & 10.25        &  9.71 &       29.282 &      -10.217 &      0.594934 &	AQ Tuc		&      (3)                          \\ 
10   &     3.926140119   &   -72.40455559  & 4689606693809057920        & 12.52        & 12.85        & 13.08        & 12.47 &       12.182 &        0.874 &      4.183775 &                    &                                       \\ 
11   &     5.617000506   &   -71.35597297  & 4689919401777839872        & 15.25        & 15.57        & 15.80        & 15.19 &        2.629 &       -2.465 &      0.271553 &	6237		&      (2)                          \\ 
12   &     4.438502576   &   -71.47351863  & 4689892051426793088        & 15.93        & 16.64        & 17.27        & 15.85 &        1.607 &        2.576 &      0.301928 &	6202		&      (2)                          \\ 
13   &     8.383993384   &   -72.44371408  & 4689170393848047616        & 16.37        & 16.34        & 16.29        & 16.28 &       -2.151 &       -0.490 &      9.700781 &	0190     	&      (2)                          \\ 
14   &     8.667676514   &   -72.42093551  & 4689175444729490176        & 16.09        & 16.58        & 16.96        & 16.02 &        7.916 &       -2.849 &      6.348956 &	6320		&      (2)                          \\ 
15   &     3.233508991   &   -72.31701594  & 4689705065728365952        & 13.87        & 14.58        & 15.23        & 13.80 &       14.594 &      -11.702 &      6.221432 &			&                                   \\ 
16   &     8.890463344   &   -72.00424710  & 4689948607555229440        & 13.31        & 14.00        & 14.62        & 13.25 &       21.512 &       -7.481 &      8.311906 &                    &                                       \\ 
17   &     9.001923031   &   -72.13692603  & 4689195235938665984        & 16.49        & 16.95        & 17.30        & 16.42 &       10.471 &       -0.916 &      2.195023 &                    &                                       \\ 
18   &     3.106567298   &   -71.63696759  & 4701807184057132544        & 15.34        & 16.07        & 16.76        & 15.26 &       19.053 &       -9.645 &      0.797952 &                    &                                       \\ 
19   &     7.236826679   &   -73.04733569  & 4688719937672536576        & 15.52        & 16.02        & 16.39        & 15.44 &        9.457 &        1.928 &      0.887469 &	0110     	&      (2)                          \\ 
20   &     6.095619143   &   -71.04839847  & 4701937201307042304        & 10.84        & 11.28        & 11.58        & 10.75 &       -9.681 &      -32.364 &      0.337970 &                    &                                       \\ 
21   &     8.348316851   &   -71.26629784  & 4690231525639747072        &  5.84        &  6.04        &  6.22        &  5.83 &       73.596 &      -12.372 &      7.108649 &	Theta Tuc	&      (3)                          \\ 
22   &     4.674966177   &   -71.02474137  & 4701931566310065280        & 16.15        & 16.73        & 17.19        & 16.06 &       17.406 &        1.644 &      3.666514 &	6209  		&      (2)                          \\ 
23   &     2.505425028   &   -71.58902403  & 4701817938655245568        & 10.51        & 10.77        & 10.95        & 10.48 &        2.737 &        3.244 &      2.727260 &                    &                                       \\ 
24   &     9.268095200   &   -71.43438559  & 4690047048205684736        & 15.86        & 16.11        & 16.27        & 15.80 &        4.118 &        4.035 &      0.462172 &	0301      	&      (2)                          \\ 
25   &     2.113365648   &   -72.31207417  & 4689760969022718720        & 13.08        & 13.49        & 13.79        & 13.02 &       -0.194 &        0.552 &      0.330047 &	6174		&      (2)                          \\ 
26   &     4.262175365   &   -70.99149906  & 4701976680646546688        & 11.58        & 12.16        & 12.65        & 11.52 &       15.144 &      -10.681 &      1.406333 &                    &                                       \\ 
27   &     2.826367562   &   -71.30014165  & 4701877209203898368        & 13.83        & 14.37        & 14.83        & 13.76 &      -12.364 &       -6.930 &      0.269175 &	6181            &                                   \\ 
28   &     4.266066198   &   -70.91246927  & 4701989595612092416        & 15.96        & 16.92        & 18.00        & 15.88 &        6.717 &       -7.449 &      0.302563 &                    &                                       \\ 
29   &     5.461765066   &   -70.77851842  & 4702006058222771072        & 12.30        & 12.82        & 13.27        & 12.25 &       11.954 &       -7.653 &      0.271736 &                    &                                       \\ 
30   &     2.157617319   &   -71.48742290  & 4701865698691623680        & 13.59        & 14.03        & 14.39        & 13.54 &        2.934 &        1.175 &      0.305206 &	6175		&      (2)                          \\ 
31   &     7.088826924   &   -73.39119841  & 4688521407109977344        & 15.68        & 16.48        & 17.28        & 15.61 &        0.379 &       -1.267 &      1.705885 &	         	&                                   \\ 
32   &     5.850347702   &   -73.53212906  & 4688493335202967680        & 16.44        & 16.84        & 17.10        & 16.34 &        0.714 &       -4.793 &      0.385427 &	0024      	&      (2)                          \\ 
33   &     2.534043926   &   -71.11817683  & 4701889887947453824        & 15.58        & 16.01        & 16.34        & 15.51 &       19.158 &       -8.877 &      0.177335 &                    &                                     \\ 
34   &     3.200471157   &   -73.29941528  & 4688666370837766144        & 14.62        & 14.92        & 15.12        & 14.56 &        2.890 &      -10.635 &      1.917045 &                    &                                     \\ 
  \hline
\multicolumn{13}{l}{{\it References} (1) \citet[updated to 2017]{2001AJ....122.2587C}; (2) \citet{2016AcA....66..421P}; (3) \citet{2017ARep...61...80S}
}
\end{tabular}

      }
\end{table*}

\section{Variable stars}
\label{sec:varsta}
Variable star detection has been performed using 
three different algorithms: the Generalized Lomb-Scargle
(GLS) periodogram (\citealt{2009A&A...496..577Z}),  the
Analysis of Variance (AoV) periodogram
(\citealt{1989MNRAS.241..153S}), and the BLS periodogram. The
procedure is illustrated in Fig.~\ref{fig11}. We isolated the
candidate variable stars using the Signal-to-Noise Ratio (SNR)
parameter and adopting the procedure described in
\citet{2015MNRAS.447.3536N}: first, we produced the histograms of the
detected periods for all the light curves and suppressed the spikes
due to systematic effects in the light curves. We divided the SNR
distributions in bins of period $\delta P =1$\,d and we computed the
3.5$\sigma$-clipped mean and standard deviation values of the SNR
within each period interval. We interpolated the points 4$\sigma$
above the mean values with a spline and selected all the stars above
this line (orange points in panels (a),  (b), and
  (c) of Fig.~\ref{fig11}). The number of candidate variable stars so
selected are  2446 and 561 according the
GLS and AoV algorithms, respectively. The variables identified by both
the techniques are  502. The number of variable
  sources found with BLS periodogram is 391.  We used a routine of
  \texttt{VARTOOLS} (\citealt{2016A&C....17....1H}) to exclude the
  sources that show variability because blended with a real variable
  star. After excluding blends and after a visual inspection, we
  identified 34 eclipsing binaries.  
Positions of the candidate variables in the VPD and CMD are
plotted in blue (field stars), red (47\,Tuc members), and green (SMC
stars) in panels  (d) and (e) of
Fig.~\ref{fig11}, respectively. Panels  (f) show
some examples of variable stars, colour-coded as in panels (d) and (e). The magnitude of these stars spans from $T\sim 7.4$ to $T\sim
13.4 $.

\subsection{Eclipsing Binaries}

  We detected 34 eclipsing binaries. Their phased light curves
  are shown in Fig.~\ref{fig12} colour-coded as the variables in
  Fig.~\ref{fig11}.  Among the detected eclipsing binaries, 4 of them
  have a high probability to be 47\,Tuc members (in red), one is a
  star in the SMC (in green), and the other 28 stars belong to the
  Galactic field (in blue). In Table~\ref{tab2} we listed the 34
  eclipsing binaries.  We cross-identified our list of eclipsing
  binaries with other catalogues in literature. We found 6 stars in
  common with the catalogue by \citet[updated to
    2017]{2001AJ....122.2587C}, 13 listed also in the catalogue of
  eclipsing binaries based on OGLE data
  (\citealt{2016AcA....66..421P}), and two in the GCVS catalogue
  (\citealt{2017ARep...61...80S}). The period distribution of
  the detected eclipsing binaries is peaked at $\sim$0.3-0.4\,d,
  and $\sim 58$\,\% of stars have P$<1$\,d and are contact
  binaries. 

\subsection{RGB and AGB stars of 47\,Tuc}
\begin{figure*}
  \includegraphics[width=0.995\textwidth,bb= 15 300 590  688]{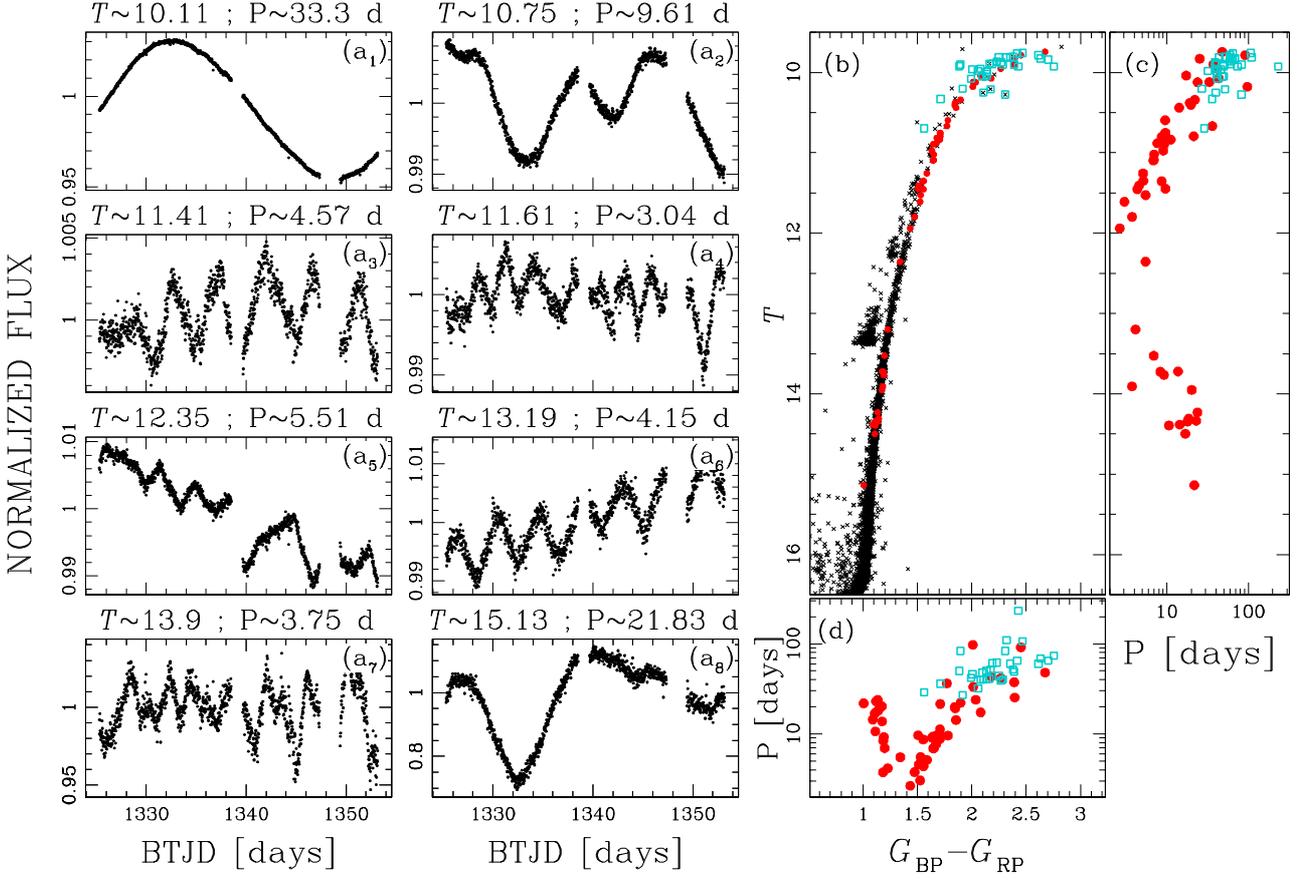}
  \caption{Analysis of RGB stars. Panels (a$_1$)-(a$_8$) show a sample
    of light curves of RGB stars, ordered by magnitude. Panel (b)
    shows $T$ versus $G_{\rm BP}-G_{\rm RP}$ CMD; panels (c) and (d) are
    the PL and the PC relations. In panels (b), (c), and (d) azure
    squares are determined by \citet{2005A&A...432..207L}, red points
    are obtained in this work.  \label{fig13}}
\end{figure*}

  We analysed the RGB stars members of 47\,Tuc, that we flagged
  as candidate variable, to find their variability periods.  To
  minimise the effects of the blends, we excluded from the analysis
  the RGB stars located at a distance $<7.5$\,arcmin from the cluster
  centre.  We also excluded all the stars that are blends, by comparing
  their light curves (and the periods found by GLS/AoV) with those of
  the neighbours located at a distance of 100\,arcsec. After these
  selections, we saved 54 RGB stars with a good light curve. A sample
  of these light curves is shown in panels (a) of Fig.~\ref{fig13},
  while in panel (b) we show their position (red points)
  on the $T$ versus $G_{\rm BP}-G_{\rm RP}$ CMD. We re-extracted the
  GLS periodograms for these light curves, but in this second
  iteration we looked for periods between 0.1 and 150 days. We listed
  the 54 RGB stars, their astro-photometric properties and their
  periods in Table~\ref{tab3}.
   
  In this paper we extend the work by
  \citet{2005A&A...432..207L}: they analysed the long period,
  pulsating stars located on the tip of the RGB in the
  Period-Luminosity (PL) plan. Because of the uncertainties on their
  photometric series, their study was limited to stars with $V\lesssim
  12$. Thanks to the high quality of our light curves, we are able to
  extend the PL distribution to fainter magnitudes ($T\sim
  15.1$). Panel (c) of Fig.~\ref{fig13} shows the relation between $T$
  magnitude and the period for the RGB stars analysed by \citet[azure
    squares]{2005A&A...432..207L} and by us (red points)\footnote{Only
    one star is in common to both the samples, and the period found by
    \citet{2005A&A...432..207L} and by us differs of $\sim 1$\,day.}: the
  two distributions are superimposed. In the range $9.5 \lesssim T
  \lesssim 12.0$, the period is proportional to the luminosity of the
  stars. In the range $12.0 \lesssim T \lesssim 15.5$ we found the
  unexpected result that the period is inversely proportional to the
  luminosity. For completeness, we show the relation Period-Colour
  (PC) in panel (d) of Fig.~\ref{fig13}: for $(G_{\rm BP}-G_{\rm
    RP})\gtrsim 1.5$ the period grows with the colour, while for
  $(G_{\rm BP}-G_{\rm RP})\lesssim 1.5$ the period decreases with the
  colour.  

\begin{figure*}
  \includegraphics[width=0.7\textwidth,bb= 15 385 465  688]{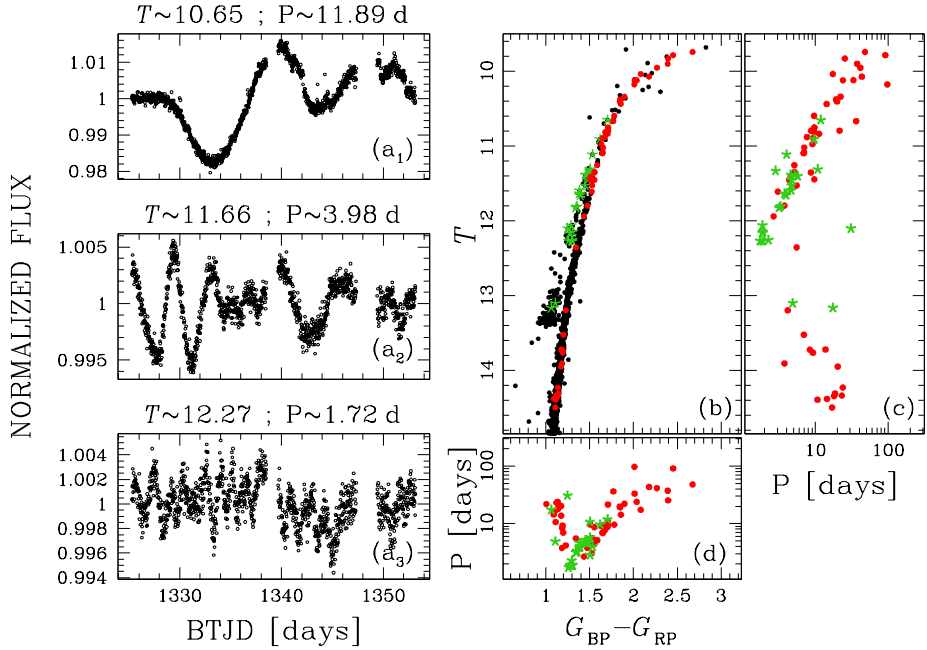}
  \caption{Analysis of AGB stars. Panels (a$_1$)-(a$_3$) show a sample
    of light curves of AGB stars, ordered by magnitude. Panel (b)
    shows $T$ versus $G_{\rm BP}-G_{\rm RP}$ CMD; panel (c) and (d)
    the PL and the PC relations. In panels (b), (c), and (d) red circles and green starred points  
    are RGB and AGB stars, respectively.  \label{fig14}}
\end{figure*}

 We did the same analysis with the asymptotic giant-branch (AGB)
  stars members of 47\,Tuc. We found 24 AGB variable stars with good
  light curves; they are listed in Table~\ref{tab4}. We derived the period by using the GLS periodogram
  and looking for periods between 0.1 and 150 days. Finally we analysed
  the PL and PC distributions. Figure~\ref{fig14} summarises the
  results. Panels (c) and (d) show that the PL and PC distributions
  for AGB (in green) and RGB (in red) stars are superimposed. 

 A detailed analysis of the PL/PC distributions for AGB and RGB
  stars in globular clusters is beyond the scope of this work
  and might be the subject of a future analysis based on our released light curves.

\begin{table*}
  \caption{List of the analysed RGB stars in 47\,Tuc}
    \label{tab3}
    \resizebox{0.8\textwidth}{!}{
      \begin{tabular}{c c c c c c c c c }
\hline
{\bf $\alpha$(J2000)} & {\bf $\delta$(J2000)} & {\bf Gaia DR\,2 ID} & {\bf $T$} & {\bf $G$} & {\bf $G_{\rm BP}$} & {\bf $G_{\rm RP}$} & {\bf P} & {\bf $\sigma_P$}  \\
{($^\circ$)} & {($^\circ$)} & {} & {} & {} & {} & {} & {(d)} & {(d)} \\
\hline
\multicolumn{9}{c}{RGB} \\
\hline
6.42920058   &   $-72.06821514$ &  4689627069134206080  &  13.72 & 14.32 & 14.82 &   13.65   &      13.76     &       0.10    \\    
6.43662601   &   $-72.08778119$ &  4689627000414752896  &  14.31 & 14.89 & 15.38 &   14.23   &      18.62     &       0.26    \\
5.66859934   &   $-72.14857213$ &  4689637407107299840  &  10.97 & 11.76 & 12.55 &   10.91   &       9.08     &       0.13    \\
6.01883235   &   $-71.95310074$ &  4689644730042501760  &  10.60 & 11.43 & 12.32 &   10.54   &       9.59     &       0.06    \\
6.12140329   &   $-72.20647551$ &  4689623010398337408  &  13.95 & 14.54 & 15.05 &   13.88   &      20.27     &       0.21    \\
5.64847070   &   $-72.02281411$ &  4689642698537939328  &  13.72 & 14.32 & 14.84 &   13.65   &       8.37     &       0.11    \\
6.43320055   &   $-72.11410571$ &  4689626725536883328  &   9.95 & 10.96 & 12.17 &    9.90   &      41.55     &       ---     \\
6.44980872   &   $-72.08972115$ &  4689626794256322048  &  14.39 & 14.96 & 15.42 &   14.31   &      10.65     &       0.17    \\
6.45352017   &   $-72.08855492$ &  4689626794256321024  &  14.35 & 14.92 & 15.36 &   14.23   &      18.00     &       0.18    \\
6.41109649   &   $-72.01835333$ &  4689639919678548736  &  11.41 & 12.15 & 12.82 &   11.31   &       4.58     &       0.02    \\
5.59606753   &   $-72.11807831$ &  4689638373492005888  &  10.75 & 11.56 & 12.40 &   10.69   &       9.66     &       0.14    \\
6.37838253   &   $-71.99910431$ &  4689640194556564224  &  14.38 & 14.94 & 15.39 &   14.30   &      14.44     &       0.12    \\
5.57456597   &   $-72.10343371$ &  4689638442211439872  &  10.12 & 11.04 & 12.08 &   10.07   &      33.31     &       ---     \\
6.35701402   &   $-72.17760343$ &  4689623289563167360  &  10.12 & 11.05 & 12.11 &   10.07   &      23.89     &       0.34    \\
6.49858087   &   $-72.07526996$ &  4689632875931941888  &  11.02 & 11.81 & 12.61 &   10.96   &       7.02     &       0.07    \\
6.48973686   &   $-72.13133662$ &  4689629371238695552  &  11.61 & 12.35 & 13.07 &   11.55   &       3.04     &       0.01    \\
6.51160277   &   $-72.05507897$ &  4689632974708827776  &  13.20 & 13.82 & 14.36 &   13.13   &       4.16     &       0.03    \\
6.34367313   &   $-72.19870524$ &  4689622460621134720  &  11.35 & 12.11 & 12.85 &   11.29   &       8.67     &       0.15    \\
5.83524393   &   $-71.93883381$ &  4689831200329811840  &  14.50 & 15.06 & 15.53 &   14.43   &      16.91     &       0.24    \\
5.52387023   &   $-72.06572626$ &  4689618273040536704  &  10.67 & 11.50 & 12.38 &   10.61   &      36.33     &       ---     \\
5.64934543   &   $-72.18648647$ &  4689625282428885120  &   9.78 & 10.84 & 12.21 &    9.76   &      91.28     &       ---     \\
6.03188484   &   $-71.92610741$ &  4689645004920347264  &  15.14 & 15.65 & 16.07 &   15.06   &      21.84     &       0.38    \\
6.21701197   &   $-71.93645739$ &  4689644420804776960  &   9.90 & 10.94 & 12.25 &   9.865   &      37.41     &       ---     \\
6.21656039   &   $-72.22729200$ &  4689622052612663680  &  14.23 & 14.81 & 15.29 &   14.16   &      23.82     &       0.42    \\
6.31562558   &   $-72.21235971$ &  4689622430569760640  &  10.89 & 11.68 & 12.48 &   10.83   &       9.34     &       0.07    \\
5.52469338   &   $-72.13027478$ &  4689614901478662272  &  10.40 & 11.27 & 12.19 &   10.34   &      19.83     &       0.12    \\
6.05234523   &   $-71.92107173$ &  4689645107999550720  &  11.80 & 12.52 & 13.21 &   11.73   &       3.77     &       0.03    \\
6.45510154   &   $-72.19421879$ &  4689622396209979136  &  14.34 & 14.91 & 15.37 &   14.25   &      23.03     &       1.14    \\
6.28343260   &   $-71.92632578$ &  4689644352084823936  &  13.77 & 14.37 & 14.89 &   13.70   &       9.25     &       0.09    \\
5.98279786   &   $-71.90490959$ &  4689832712158218880  &  12.36 & 13.02 & 13.64 &   12.29   &       5.51     &       0.05    \\
6.51617594   &   $-71.97513601$ &  4689645898275596288  &  13.53 & 14.13 & 14.66 &   13.46   &       6.95     &       0.06    \\
5.71571955   &   $-72.24380077$ &  4689624212967816064  &  11.09 & 11.88 & 12.67 &   11.02   &       6.96     &       0.05    \\
6.09369763   &   $-71.89129882$ &  4689832849597140096  &   9.83 & 10.87 & 12.19 &    9.79   &      25.36     &       2.28    \\
6.40041875   &   $-72.23750838$ &  4689575289007212416  &  11.45 & 12.18 & 12.89 &   11.38   &       9.65     &       0.08    \\
5.91328145   &   $-72.27775711$ &  4689620334626390784  &  10.38 & 11.24 & 12.17 &   10.32   &      19.23     &       0.03    \\
5.39619011   &   $-71.99971800$ &  4689807354671547648  &  10.04 & 10.98 & 12.08 &    9.99   &      17.35     &       0.28    \\
5.57486258   &   $-72.26169193$ &  4689600680854700288  &  10.79 & 11.61 & 12.44 &   10.73   &      21.47     &       0.21    \\
5.68279610   &   $-71.87548264$ &  4689835082980178304  &  11.94 & 12.65 & 13.31 &   11.88   &       2.66     &       0.01    \\
6.76144736   &   $-72.01953661$ &  4689633460047351296  &  11.35 & 12.10 & 12.84 &   11.28   &       5.17     &       0.03    \\
6.51675254   &   $-71.88983356$ &  4689650330681896320  &  10.34 & 11.22 & 12.18 &   10.28   &      22.14     &       0.05    \\
5.18998805   &   $-72.07752865$ &  4689806457019852160  &  10.82 & 11.62 & 12.43 &   10.76   &       9.63     &       0.06    \\
6.25387493   &   $-72.34392027$ &  4689573364861989760  &  11.10 & 11.89 & 12.68 &   11.04   &       6.88     &       0.04    \\
5.31194405   &   $-72.25063322$ &  4689612981641053696  &  10.76 & 11.57 & 12.41 &   10.70   &       9.61     &       0.06    \\
5.22378507   &   $-72.22670400$ &  4689613183491785856  &  11.45 & 12.21 & 12.95 &   11.39   &       4.38     &       0.02    \\
5.97192517   &   $-71.78663649$ &  4689837419442795648  &   10.8 & 11.61 & 12.44 &   10.74   &       8.72     &       0.05    \\
5.87516442   &   $-72.37674822$ &  4689572677667393664  &   9.74 & 10.86 & 12.39 &    9.71   &      47.97     &       ---     \\
6.61872853   &   $-71.83689944$ &  4689651086596023040  &  11.26 & 12.02 & 12.78 &   11.19   &       5.13     &       0.03    \\
5.68086357   &   $-71.77517546$ &  4689848655076717440  &  11.53 & 12.27 & 12.99 &   11.47   &       5.53     &       0.05    \\
7.06600513   &   $-72.15084334$ &  4689582435832666752  &  10.07 & 11.05 & 12.18 &   10.00   &      43.35     &       ---     \\
6.61344377   &   $-71.79558420$ &  4689652117388263296  &  10.88 & 11.68 & 12.49 &   10.82   &       7.67     &       0.08    \\
6.73945779   &   $-71.82054282$ &  4689651151009833728  &  10.84 & 11.65 & 12.48 &   10.77   &      11.22     &       0.15    \\
5.41339391   &   $-72.41406304$ &  4689595831823970304  &  10.18 & 11.10 & 12.13 &   10.12   &      97.15     &       ---     \\
6.62113569   &   $-71.66616431$ &  4689844287094705408  &  13.91 & 14.51 & 15.03 &   13.84   &       3.76     &       0.01    \\
6.79559874   &   $-72.51910046$ &  4689558074778482304  &  10.44 & 11.30 & 12.24 &   10.38   &      14.25     &       0.05    \\
\hline
\end{tabular}
 
      }
\end{table*}

\begin{table*}
  \caption{List of the analysed AGB stars in 47\,Tuc}
  \label{tab4}
    \resizebox{0.8\textwidth}{!}{
      \begin{tabular}{c c c c c c c c c }
\hline
{\bf $\alpha$(J2000)} & {\bf $\delta$(J2000)} & {\bf Gaia DR\,2 ID} & {\bf $T$} & {\bf $G$} & {\bf $G_{\rm BP}$} & {\bf $G_{\rm RP}$} & {\bf P} & {\bf $\sigma_P$}  \\
{($^\circ$)} & {($^\circ$)} & {} & {} & {} & {} & {} & {(d)} & {(d)} \\
\hline
\multicolumn{9}{c}{AGB} \\
\hline
6.01688895   &     -71.94061171     &      4689644798761947520  & 10.65  & 11.47  & 12.30  & 10.60 &    11.89 &     0.25  \\
5.55209154   &     -72.09583082     &      4689641392838717568  & 12.10  & 12.73  & 13.29  & 12.04 &    30.93 &     0.08  \\
6.47697044   &     -72.03029288     &      4689633975443492736  & 11.39  & 12.10  & 12.78  & 11.33 &     5.15 &     0.04  \\
5.54159063   &     -72.05543971     &      4689642496660354048  & 11.45  & 12.18  & 12.88  & 11.39 &     4.74 &     0.07  \\
5.96855292   &     -71.93058778     &      4689645039280106624  & 13.17  & 13.71  & 14.17  & 13.10 &    17.40 &     0.19  \\
6.34863462   &     -72.20215239     &      4689622460621136256  & 11.83  & 12.50  & 13.11  & 11.77 &     3.14 &     0.03  \\
5.84093194   &     -71.93252052     &      4689831264750937344  & 11.40  & 12.14  & 12.85  & 11.35 &     5.70 &     0.06  \\
6.39289277   &     -71.95979283     &      4689641083606519168  & 11.64  & 12.32  & 12.96  & 11.58 &     3.78 &     0.04  \\
5.47065091   &     -72.14455881     &      4689614867118928384  & 11.38  & 12.11  & 12.80  & 11.33 &     4.59 &     0.03  \\
5.81541011   &     -71.89896796     &      4689832192474832000  & 11.11  & 11.86  & 12.59  & 11.06 &     4.00 &     0.04  \\
6.34588623   &     -71.91017173     &      4689644592602980608  & 11.33  & 12.07  & 12.78  & 11.27 &     2.85 &     0.03  \\
5.39536304   &     -72.14050345     &      4689617792004292096  & 12.06  & 12.71  & 13.29  & 12.00 &     1.87 &     0.02  \\
5.50575090   &     -72.22177206     &      4689601746006553344  & 11.31  & 12.05  & 12.76  & 11.25 &    10.75 &     0.17  \\
6.76078109   &     -72.02121070     &      4689633253888922624  & 13.10  & 13.67  & 14.15  & 13.04 &     4.88 &     0.02  \\
6.59880135   &     -71.89197969     &      4689647405798359936  & 12.26  & 12.91  & 13.49  & 12.19 &     2.27 &     0.02  \\
6.72736160   &     -72.22744307     &      4689580271169229824  & 12.27  & 12.93  & 13.52  & 12.22 &     1.87 &     0.01  \\
5.55421378   &     -72.30091451     &      4689600332949678848  & 10.91  & 11.69  & 12.47  & 10.85 &     9.64 &     0.08  \\
6.43414368   &     -72.31488155     &      4689573914617758336  & 12.15  & 12.80  & 13.40  & 12.09 &     1.89 &     0.01  \\
6.95294326   &     -71.95535448     &      4689635178034205056  & 11.81  & 12.48  & 13.11  & 11.75 &     3.33 &     0.02  \\
5.25552963   &     -72.31604682     &      4689600852653460352  & 11.59  & 12.28  & 12.92  & 11.54 &     4.64 &     0.02  \\
7.13714170   &     -72.10676444     &      4689583294826108160  & 12.16  & 12.81  & 13.38  & 12.11 &     1.88 &     0.01  \\
5.41790367   &     -72.38729162     &      4689596111009591936  & 12.27  & 12.91  & 13.47  & 12.21 &     1.72 &     0.01  \\
5.30446552   &     -71.64086650     &      4689877654695832704  & 11.66  & 12.35  & 12.99  & 11.60 &     3.98 &     0.02  \\
4.10826661   &     -72.35587640     &      4689607995176575488  & 11.53  & 12.24  & 12.92  & 11.47 &     4.66 &     0.08  \\
\hline
\end{tabular}

      }
\end{table*}

\begin{figure*}
  \includegraphics[width=0.75\textwidth,bb= 21 190 565  717]{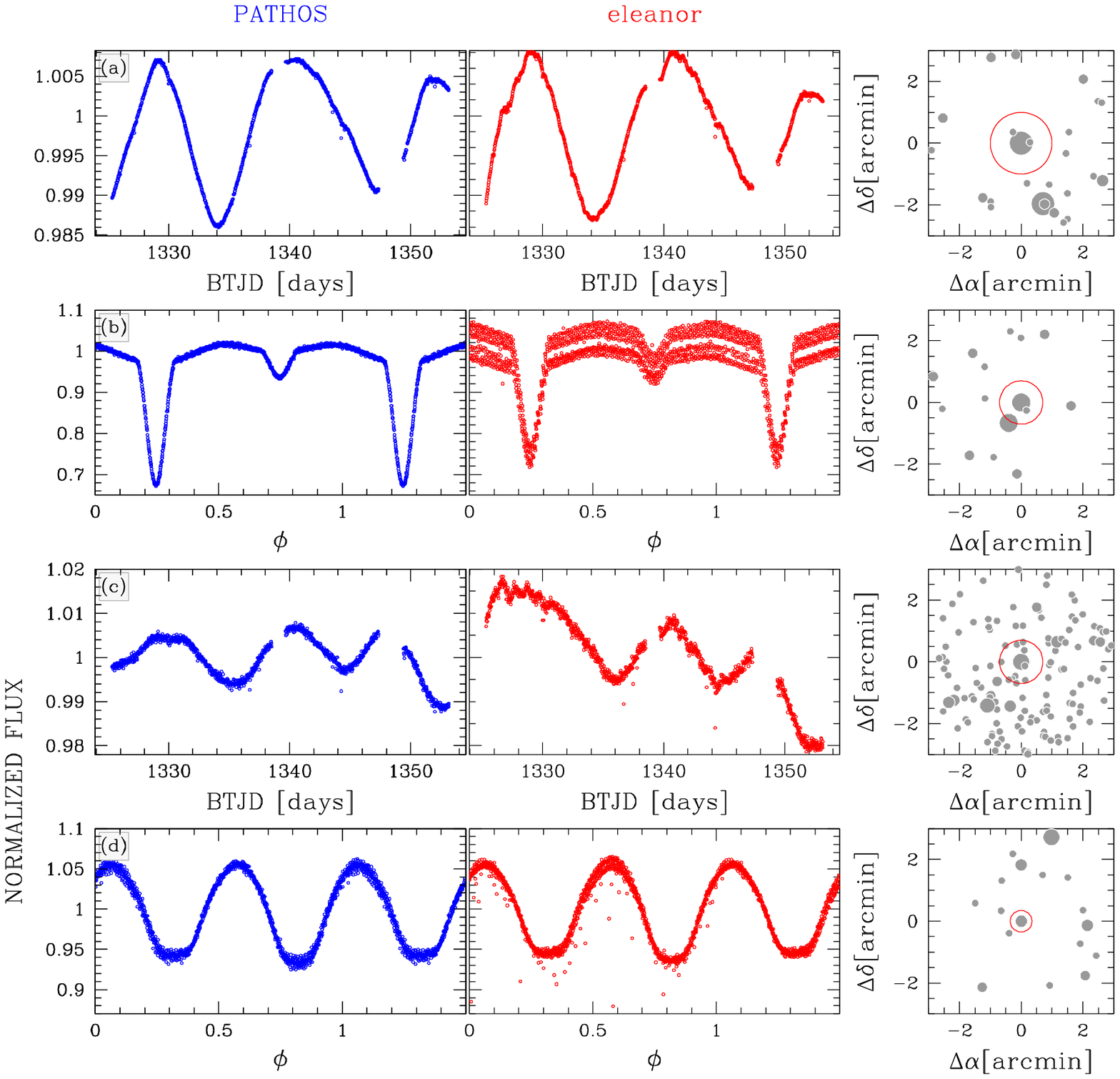} \\
  \includegraphics[width=0.75\textwidth,bb= 21 441 565  685]{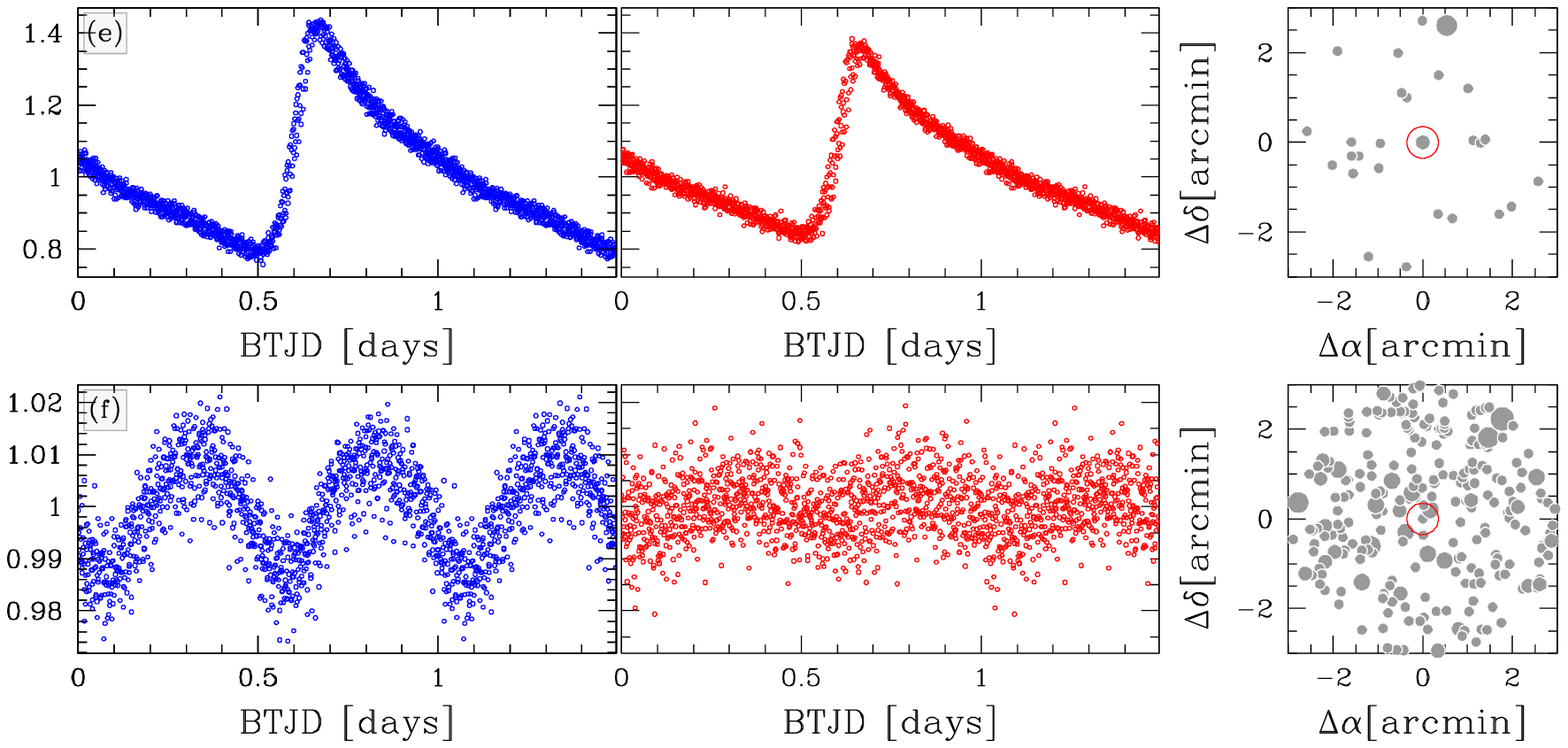}
  \caption{Comparison between the raw light curves extracted
      with the PATHOS pipeline (in blue, left panels) and the
      \texttt{eleanor} pipeline (in red, middle panels). Right panels
      show the finding charts of the target stars, obtained by using the Gaia\,DR2
      data: in red the size of the aperture used for the extraction of
      the light curves. The light curves are sorted by magnitude. \label{fig15}}
\end{figure*}

\section{Comparison with other pipelines}
\label{sec:compare}
  At present, the most advanced pipeline aimed at the
  extraction of any kind of light curve from FFIs is \texttt{eleanor}
  (\citealt{2019PASP..131i4502F}). In this section we compare the
  quality of our photometry with that obtained by \texttt{eleanor},
  highlighting the importance of neighbour subtraction when we consider
  crowded environments.  Figure~\ref{fig15} shows six different raw
  light curves of stars with different luminosities and located in
  environments characterised by different levels of crowding.  Panels
  on the right show that the contamination by neighbour stars in the
  cases (a) (Gaia\,DR2\,4689720939927488768, $T\sim 7.7$), (d)
  (Gaia\,DR2\,4689760969022718720, $T\sim 13.1$), and (e)
  (Gaia\,DR2\,4689953108681026304, $T\sim 15.6$ ) is low, and the raw
  light curves output of the PATHOS pipeline (left panels, in blue)
  and \texttt{eleanor} pipeline (middle panels, in red) are similar.
  The light curves were extracted by using the same photometric apertures. In
  crowded environments, the dilution and contamination by neighbour
  stars affect the quality of the light curves extracted with
  \texttt{eleanor}. In the case (b) (Gaia\,DR2\,4701817938655245568,
  $T\sim10.5$) the bright neighbour star dilutes the light of the
  target star during the eclipses, with the result that the observed
  depth of the eclipses is lower. This effect is important especially
  in the light curve analysis of candidate transiting
  exoplanets: without considering the dilution effects due to
  neighbour stars, the radius of the candidate exoplanet would be
  underestimated. In the cases (c) (Gaia\,DR2\,4689837419442795648,
  $T\sim10.8$) and (f) (Gaia\,DR2\,4689619853580301824, $T\sim 15.9$)
  the presence of many neighbour stars strongly affects the shape and
  the quality of the light curve of the target star; in particular, in
  the case (f) the variability of the target star is significantly diluted
  by the light of the neighbour stars.  

\section{The data release}
\label{sec:dr}
We publicly release all the light curves extracted in this work. The
light curves will be available in the Mikulski Archive for Space
Telescopes (MAST) as a High Level Science Product (HLSP) via \url{https://doi.org/10.17909/t9-es7m-vw14}.  Each light
curve contains the epoch in BTJD, the 5 extracted photometries
(PSF-fitting, 1-pixel, 2-pixel, 3-pixel, 4-pixel aperture), the value
of the local sky, the position $(x,y)$ on the image, and the data
quality flag (see Sect.~9 of the TESS Science Data Products
Description
Document\footnote{    \resizebox{0.449\textwidth}{!}{
\url{https://archive.stsci.edu/missions/tess/doc/EXP-TESS-ARC-ICD-TM-0014.pdf}}},
for details). Light curves are both in \texttt{ascii} and
\texttt{fits} format, which header contains information on the star
(from the Gaia\,DR2 catalogue) and on its observations.

\section{Summary}
\label{sec:conc}
 
In this work we  presented our PSF-based approach, applied for the
first time to {\it TESS} FFIs, in order to extract light curves of
stars in a crowded field centred on 47\,Tuc. The pipeline presented
in this pilot work is essential for the success of the PATHOS project,
whose main scope is the analysis of high-precision light curves of
stellar cluster members, in order to find candidate exoplanets
orbiting bright cluster stars and variable stars.

Discovering and characterising exoplanets in stellar clusters
(especially open clusters and young associations) represent an important ingredient to
understand how exoplanet systems have formed and evolved. At variance with what happens for most
 Galactic field stars, cluster star parameters (such as age, mass,
and chemical composition) are generally well
determined with high accuracy. This allows us to correlate stellar
parameters (such as stellar mass) with exoplanet
characteristics. Furthermore, hundreds of stellar clusters populate
the Milky Way (\citealt{2016A&A...588A.120C}), having ages that span
from few tens Myrs to $\sim 10$ Gyrs (\citealt{2019A&A...623A.108B})
and showing a wide variety of chemical compositions. Combining pieces of
information on stellar clusters with exoplanet properties (and with the
rate of exoplanets in stellar clusters), it will be possible to
understand how ages of stars, and the environment in which they are embedded,
have affected the formation and evolution of exoplanets.

In this pilot project we have extracted, corrected, and analysed the
light curves of 16641 stars in a field containing the globular cluster
47\,Tuc. These stars belong either to 47\,Tuc, or the Milky Way, or
the SMC.  We took advantage of the Gaia DR2 catalogue to extract not
only astro-photometric data of stars located in this region, but also,
when possible, stellar parameters. We searched for transit signals
among the extracted light curves and we found and characterised a
candidate transiting exoplanet orbiting a MS field star. This
candidate exoplanet, named PATHOS-1, is not in the list of {\it TESS}
Objects of Interest (TOI)\footnote{\url{ https://tev.mit.edu/toi/}}
and is a new discovery.  No candidate transiting exoplanets have been
found around bright RGB stars of 47\,Tuc; given the nature of RGB
stars, the probability of finding exoplanets orbiting them and having
periods $<27$\,days, is low. We also searched for variables among the
stars in the analysed field. 
We analysed the 34 eclipsing binaries in the field and the
  relations PL and PC for the RGB and AGB stars members of 47\,Tuc. We
  found that the period of variability for AGB and RGB stars is
  proportional to the luminosity above the mean magnitude of the RGB
  bump, but we found also that below the RGB bump the period decreases
  as the stellar brightness increases.

The raw and corrected light curves will be uploaded on the MAST
archive as HLSP. In this way, they will always be available to the
astronomical community for any scientific goal.

\section*{Acknowledgements}
PATHOS data products are available at MAST: \url{http://archive.stsci.edu/hlsp/pathos}.

DN and GP acknowledge partial support by the Universit\`a degli Studi
di Padova Progetto di Ateneo BIRD178590. LB received support from
Italian Space Agency (ASI) regulated by ``Accordo ASI-INAF
n. 2013-016-R.0 del 9 Luglio 2013 e integrazione del 9 Luglio
2015''. VG, MM, GP recognize partial support from the Italian Space
Agency (ASI), regulated by ``Accordo ASI-INAF n. 2015-019-R.0 del 29
Luglio 2015''.  LRB acknowledges support by MIUR under PRIN program
\#2017Z2HSMF. GL acknowledges support by CARIPARO Foundation,
according to the agreement CARIPARO-Universit\`a degli Studi Di
Padova, Pratica n. 2018/0098. Stacked image is obtained by using
\texttt{SWarp} (\citealt{2002ASPC..281..228B}) and
\texttt{ASTROMETRY.NET} codes (\citealt{2010AJ....139.1782L}). Some
tasks of the data reduction and analysis have been carried out using
\texttt{PARALLEL} (\citealt{Tange2011a}) and
\texttt{VARTOOLS}\,v.~1.36 (\citealt{2016A&C....17....1H}).



\bibliographystyle{mnras}
\bibliography{biblio}


\bsp	
\label{lastpage}
\end{document}